\documentclass[sigconf]{acmart}

\usepackage{amsmath,amsfonts}
\usepackage{algorithmic}
\usepackage{algorithm}
\usepackage{array}
\usepackage[caption=false,font=normalsize,labelfont=sf,textfont=sf]{subfig}
\usepackage{stfloats}
\usepackage{url}
\usepackage{verbatim}
\usepackage{graphicx}
\usepackage{hyperref}
\usepackage{xurl}

\usepackage{multirow}
\usepackage{tabularx}
\usepackage{booktabs}
\usepackage{color}
\usepackage{threeparttable}
\usepackage{subfig}
\usepackage{enumitem}
\usepackage{rotating}
\usepackage{bm}

\newcommand{\shortname}{RoleGen}
\newcommand{\longname}{Generative Recommendation Framework with Counterfactual Functional Role Reasoning}

\AtBeginDocument{%
  }

\setcopyright{acmlicensed}
\copyrightyear{2018}
\acmYear{2018}
\acmDOI{XXXXXXX.XXXXXXX}
\acmConference[Conference acronym 'XX]{Make sure to enter the correct
  conference title from your rights confirmation email}{June 03--05,
  2018}{Woodstock, NY}
\acmISBN{978-1-4503-XXXX-X/2018/06}

\begin{document}

\title{Awakening Dormant Users: Generative Recommendation with Counterfactual Functional Role Reasoning}

% 标题候选
% Awakening Dormant Users: A Generative Recommendation Framework with Counterfactual Functional Role Reasoning
% Awakening Dormant Users via Counterfactual Functional Role Reasoning: A Generative Framework
% Knowledge-Guided Functional Role Reasoning: A Generative Paradigm for Sparsity-Plagued Recommendation
% Dual-Stage Counterfactual Instruction and Knowledge-Driven Generative Recommendation for Dormant Users
% Counterfactual Reasoning for Generative Recommendation: Reviving Dormant Users through Dual-Stage Functional Role Inference
% ReGen: A Reasoning-Augmented Generative Framework for Awakening Dormant Users
% Breaking the Feedback Loop: Counterfactual Functional Role Reasoning for Dormant User Activation

% 一句话：我们提出一种“推理增强生成”框架，通过大模型构建反事实的“语义功能角色”路径，在缺乏转化信号的困境下打破反馈闭环，实现沉睡用户的有效唤醒。
% Vision Statement: We propose a Reasoning-Augmented Generative framework that leverages LLMs to infer counterfactual functional role trajectories, thereby breaking the feedback scarcity loop caused by the absence of conversion signals and effectively awakening dormant users.

% We propose an LLM-enhanced generative framework for awakening dormant users, which reasons over counterfactual trajectories to uncover how specific items act as strategic triggers for future conversions.

\author{Huishi Luo}
\email{hsluo2000@buaa.edu.cn}
\authornote{Work done during an internship at Kuaishou Technology.}
\authornote{Both authors contributed equally to this research.}
\orcid{0000-0002-3553-2280}
\affiliation{%
  \department{Institute of Artificial Intelligence}
  \institution{Beihang University}
  \city{Beijing}
  \country{China}
}

\author{Shuokai Li}
\authornotemark[2]
\email{lisk260@gmail.com}
\author{Hanchen Yang}
\email{yanghanchen03@kuaishou.com}
\affiliation{%
  \institution{Kuaishou Technology}
  \city{Beijing}
  \country{China}
}

\author{Zhongbo Sun}
\authornote{Co-corresponding authors.}
\email{szb17@tsinghua.org.cn}
\author{Haojie Ding}
\email{dinghaojie@kuaishou.com}
\author{Boheng Zhang}
\email{yangxiao16@kuaishou.com}
\affiliation{%
  \institution{Kuaishou Technology}
  \city{Beijing}
  \country{China}
}

\author{Zijia Cai}
\email{yangkaiwen@kuaishou.com}
\author{Renliang Qian}
\email{qianrenliang@kuaishou.com}
\author{Fan Yang}
\email{yangfan@kuaishou.com}
\affiliation{%
  \institution{Kuaishou Technology}
  \city{Beijing}
  \country{China}
}

\author{Tingting Gao}
\email{lisize@kuaishou.com}
\author{Chenyi Lei}
\email{leichenyi@gmail.com}
\author{Wenwu Ou}
\email{ouwenweu@gmail.com}
\affiliation{%
  \institution{Kuaishou Technology}
  \city{Beijing}
  \country{China}
}

\author{Fuzhen Zhuang}
\authornotemark[3]
\email{zhuangfuzhen@buaa.edu.cn}
\orcid{0000-0001-9170-7009}
\affiliation{%
  \department{Institute of Artificial Intelligence}
  \institution{Beihang University}
  \city{Beijing}
  \country{China}
}

\renewcommand{\shortauthors}{Luo et al.}

% 提交版
% Awakening dormant users, who remain engaged but exhibit low conversion, is a pivotal driver for incremental GMV growth in large-scale e-commerce platforms. However, existing approaches often yield suboptimal results since they typically rely on single-step estimation of an item's intrinsic value (e.g., immediate click probability). This mechanism overlooks the instrumental effect of items, where specific interactions act as triggers to shape latent intent and drive subsequent decisions along a conversion trajectory. To bridge this gap, we propose RoleGen, a novel framework that synergizes a Conversion Trajectory Reasoner with a Generative Behavioral Backbone. Specifically, the LLM-based Reasoner explicitly models the context-dependent Functional Role of items to reconstruct intent evolution. It further employs counterfactual inference to simulate diverse conversion paths, effectively mitigating interest collapse. These reasoned candidate items are integrated into the generative backbone, which is optimized via a collaborative “Reasoning–Execution–Feedback–Reflection” closed-loop strategy to ensure grounded execution. Extensive offline experiments and online A/B testing on the Kuaishou e-commerce platform demonstrate that RoleGen achieves a 6.2\% gain in Recall@1 and a 7.3\% increase in online order volume, confirming its effectiveness in activating the dormant user base.

\begin{abstract}
Awakening dormant users, who remain engaged but exhibit low conversion, is a pivotal driver for incremental GMV growth in large-scale e-commerce platforms. However, existing approaches often yield suboptimal results since they typically rely on single-step estimation of an item's intrinsic value (e.g., immediate click probability). This mechanism overlooks the \textit{instrumental effect} of items, where specific interactions act as triggers to shape latent intent and drive subsequent decisions along a conversion trajectory. To bridge this gap, we propose \textbf{\shortname{}}, a novel framework that synergizes a Conversion Trajectory Reasoner with a Generative Behavioral Backbone. Specifically, the LLM-based Reasoner explicitly models the context-dependent \textit{Functional Role} of items to reconstruct intent evolution. It further employs \textit{counterfactual inference} to simulate diverse conversion paths, effectively mitigating interest collapse. These reasoned candidate items are integrated into the generative backbone, which is optimized via a collaborative “Reasoning–Execution–Feedback–Reflection” closed-loop strategy to ensure grounded execution. Extensive offline experiments and online A/B testing on the Kuaishou e-commerce platform demonstrate that \shortname{} achieves a 6.2\% gain in Recall@1 and a 7.3\% increase in online order volume, confirming its effectiveness in activating the dormant user base.
\end{abstract}




\received{20 February 2007}
\received[revised]{12 March 2009}
\received[accepted]{5 June 2009}

\maketitle

\section{Introduction}

In large-scale e-commerce platforms such as Taobao and Amazon, a substantial fraction of daily active users remains engaged but has not made any purchases over an extended period. We refer to these users as \emph{dormant users}. Empirical statistics from the Kuaishou e-commerce platform reveal that dormant users constitute over $40\%$ of Daily Active Users (DAU) and contribute to more than $35\%$ of total Page Views (PV), yet account for less than $20\%$ of total orders.
Despite their low current monetization, their high engagement implies considerable latent conversion potential. 
Accurately modeling the latent purchase intent of dormant users and \textit{awakening} them into buyers, is therefore a key driver of incremental GMV growth. 
%Furthermore, effectively serving this vast long-tail user base is essential for ecosystem sustainability, as it promotes content diversity and mitigates the Matthew effect in recommendation systems.

Awakening dormant users, however, remains a relatively under-explored challenge in industrial recommendation. 
% Existing approaches typically rely on content-based feature enrichment~\cite{tao2022sminet,huang2023aligning}, knowledge transfer such as meta-learning~\cite{zhu2021learning} and domain adaptation~\cite{liu2023contrastive,xu2025c2lrec}, or even Large Language Models (LLMs)~\cite{zhang2025llmtreerec} to alleviate data sparsity and distribution shift between cold and active users.
% Conceptually, this problem shares significant conceptual overlap with cold-start user recommendation. 
% The most related problem is cold-start user recommendation. 
This task is often formulated as a variation of cold-start recommendation.
Traditional approaches typically rely on content-based feature enrichment~\cite{tao2022sminet,huang2023aligning}, knowledge transfer techniques such as meta-learning~\cite{zhu2021learning} and domain adaptation~\cite{liu2023contrastive,xu2025c2lrec}, or even Large Language Models (LLMs)~\cite{zhang2025llmtreerec} to alleviate data sparsity and distribution shift between cold and active users.

\begin{figure}[!t]
    \centering
    \vspace{-12pt}
    \includegraphics[width=0.5\textwidth, trim=10 5 0 0, clip]{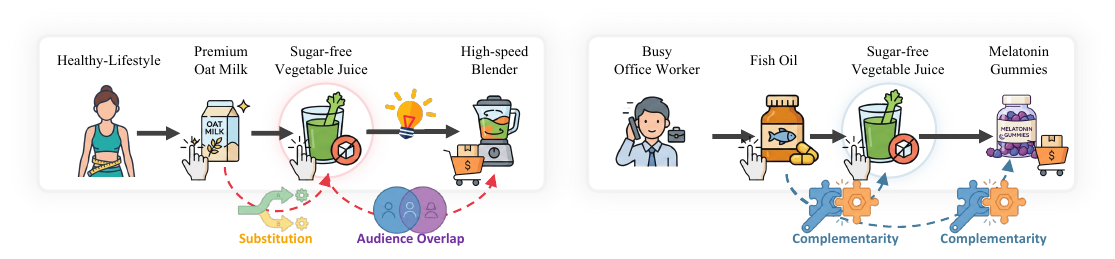}
    \vspace{-18pt}
    \caption{Instrumental effect in conversion trajectories. An item's utility extends beyond its immediate conversion probability. Instead, its contribution to the final conversion dynamically depends on the context and user intent. }
    \vspace{-15pt}
    \label{fig:intro}
\end{figure}

\textls[-12]{However, directly applying these methods to awaken dormant users often yields suboptimal results. 
Fundamentally, they model the \emph{single-step, point-wise effect} of recommending individual items, focusing solely on predicting immediate responses. 
Yet, this fails to capture real-world consumption journeys, where an item’s utility extends beyond its immediate conversion probability, i.e., \textit{intrinsic value}. Instead, items often exert an \emph{instrumental effect} on subsequent decisions: even if item $A$ is not purchased, it may still act as a trigger, shaping the user’s latent intent and facilitating conversions on other items.
Consider the example in Figure~\ref{fig:intro}: a user pursuing a healthy lifestyle clicks on ``Premium Oat Milk'' and ``Sugar-free Vegetable Juice'', and eventually purchases a “High-speed Blender”. Here, although the vegetable juice does not convert, it triggers a shift in the user’s underlying intent: freshly blended juice at home is healthier and more cost-effective than packaged drinks, which directly motivates the purchase of the blender.
Therefore, we argue that \textbf{it is essential to model an item’s instrumental effect along the user’s conversion trajectory}, rather than merely maximizing the standalone click-through rate (CTR) of individual items.
This perspective is particularly vital for dormant users with weak purchase intentions, for whom conversions often require multiple items to act in concert throughout the intent evolution, rather than any single item acting in isolation.}

Identifying these latent instrumental items poses a significant challenge.
While traditional approaches like attribution analysis~\cite{yao2022causalmta, zeng2025cabb, bencina2025lidda} may offer a solution, the extreme sparsity and noise inherent in dormant user behaviors render them ineffective. 
%The statistical co-occurrence patterns in dormant user data are too weak to support reliable discovery of implicit causal relations between items.
Given the reasoning capabilities of LLMs, leveraging their inherent world knowledge offers a promising avenue to infer latent user intents from such sparse signals. Existing methods, such as RecGPT~\cite{yi2025recgpt} and Align3GR~\cite{ye2025align3gr}, have explored integrating LLMs into recommendation. However, these works primarily utilize LLMs to infer static user profiles~\cite{liu2025onerecthink} or enrich item-level tags~\cite{yi2025recgpt}. 
Fundamentally, they still focus on evaluating the intrinsic value of items in isolation and lack explicit modeling of items’ instrumental effects on downstream conversions, which dynamically depend on the interplay among items, user intents, and consumption context.

\textls[-15]{To bridge this gap, we propose the \longname{} (\shortname{}), a dual-module framework that consists of an LLM-based \textbf{Conversion Trajectory Reasoner}, and a \textbf{Generative Behavioral Backbone} as the execution engine.
(1) The Reasoner, fine-tuned from a general LLM, explicitly models the \textbf{Functional Role} of items, characterizing how each item dynamically influences the user’s intent within a specific context.
% As illustrated in Figure~\ref{fig:intro}, the same sugar-free vegetable juice may act as a ``scenario-level substitute'' for breakfast oat milk in the context of weight loss, while serving as a ``complementary supplement'' to fish oil for a busy office worker.
% Specifically, to capture this nuance, the Reasoner performs structured reasoning over the user’s historical interactions to identify key items and their functional roles, and to construct an explicit conversion trajectory of intent evolution.
Crucially, it employs \textbf{counterfactual inference} to explore potential conversion paths, effectively mitigating the Matthew effect~\cite{yi2025recgpt} prevalent among dormant users. 
Ultimately, the Reasoner generates candidate next items represented by Semantic Identifiers (SIDs).
(2) The Behavioral Backbone serves as the execution engine trained purely on interaction logs to mitigate the potential semantic bias in LLM reasoning.
% (2) Recognizing that practical recommendation remains fundamentally driven by collaborative filtering (CF) signals, we introduce a recommendation backbone trained purely on interaction logs to mitigate the potential semantic bias in LLM reasoning.
We instantiate it as a decoder-only generative model, combining robust sequence modeling capabilities with high inference efficiency, and the ability to directly incorporate the Reasoner’s SID-level outputs as guidance.
(3) We further introduce a closed-loop collaborative training strategy: 
Reasoner counterfactually infers Functional Roles $\rightarrow$ Backbone executes recommendation $\rightarrow$ system collects feedback $\rightarrow$ Reasoner reflects and refines. This dual-module synergy enables the system to progressively capture the underlying logic of user intent evolution and substantially improve dormant user conversions.}
% todo 马太效应参考文献

In summary, our main contributions are:
\begin{itemize}[leftmargin=*, nosep]
\item We propose \shortname{}, a unified framework that synergizes an LLM-based reasoner with a generative backbone via a collaborative ``Reasoning–Execution–Feedback–Reflection'' closed-loop mechanism.
\item We introduce Functional Role Reasoning to capture the context-dependent instrumental effects of items. The Reasoner further performs counterfactual trajectory inference to simulate alternative conversion paths, thereby enhancing the diversity of generated candidate items.
\item Extensive experiments on Kuaishou demonstrate significant performance gains (offline Recall@1 +$6.2\%$ and online order volume +$7.3\%$). Moreover, \shortname{} effectively mitigates the Matthew effect by enhancing long-tail item exposure.
\end{itemize}

\section{Related Work}

The full literature review is provided in Appendix~\ref{app:related_work}.

\noindent
\textbf{Generative Recommendation}
Generative recommendation reformulates user-item interactions as a sequence generation task~\cite{wang2025generative}. Recent advances, such as TIGER~\cite{rajput2023recommender}, TokenRec~\cite{qu2025tokenrec}, and OneRec~\cite{deng2025onerec}, utilize \emph{Semantic IDs (SIDs)} to encode items into discrete tokens. By transforming recommendation into structured token prediction, these methods improve scalability and transferability. Our work builds upon this paradigm, employing a generative backbone to execute recommendations based on structured reasoning signals.
\noindent
\textbf{LLM-Enhanced Recommendation}
\textls[-16]{LLMs are increasingly leveraged to alleviate data sparsity~\cite{lin2025can,wu2024survey}. Existing approaches primarily fall into two categories: (1) {Representation Enhancement}, using LLMs to generate semantic embeddings as side information~\cite{liu2025llm-esr,liu2025llmal}; and (2) {Data Augmentation}, synthesizing pseudo-interactions to enrich training data~\cite{sun2025llmser,Wei2024LLMRec}. However, these methods typically treat LLMs as static feature extractors or isolated data generators. They lack the capability to perform explicit reasoning over user intent evolution or to tightly couple such reasoning with the execution of recommendations.}

\noindent
\textbf{Recommendation for Dormant Users}
Re-engaging dormant users is challenging due to sparse feedback and temporal preference drift~\cite{zhang2025cold,kutlimuratov2022modeling}. While traditional methods focus on latent preference recovery~\cite{chang2023latent} or temporal continuity~\cite{lee2025capturinguserinterestsdata}, they struggle when recent signals are noisy. Recent LLM-based works, such as ColdLLM~\cite{huang2024large} and CSRec~\cite{yang2024common}, explore behavioral simulation or knowledge graph augmentation to address sparsity. Yet, they predominantly focus on static profile completion or item-level cold starts. Importantly, they overlook the \textit{instrumental effect} of items in driving conversion trajectories, a gap we addresses through functional role reasoning.

\section{Method}

\subsection{Problem Formulation}
\label{sec:problem}

Let $\mathcal{U}$ and $\mathcal{I}$ denote the sets of users and items, respectively. 
For each user $u \in \mathcal{U}$, we define their historical interaction sequence as $S_u = [(i_1, b_1, t_1), (i_2, b_2, t_2), \dots, (i_{N_u}, b_{N_u}, t_{N_u})]$, ordered by timestamp $t_1 < t_2 < \dots < t_{N_u}$. 
Here, each interaction is a tuple consisting of the item $i_k \in \mathcal{I}$, the behavior type $b_k \in \mathcal{B} = \{\texttt{view}, \texttt{click},$ $\texttt{purchase}\}$, and the timestamp $t_k$.

\begin{definition}[Dormant Users]
We define a subset of users $\mathcal{U}_0 \subset \mathcal{U}$ as \textit{dormant users}. 
A user $u$ is categorized as dormant at the current time $t_{now}$ if they have not executed any purchase actions within the past $\Delta T = 30$ days, formalized as:
\begin{equation}
\label{eq:u0}
\begin{split}
\mathcal{U}_0 
= \big\{ u \in \mathcal{U} \,\big|\,
  \nexists (i_k, b_k, t_k) \in S_u :\; b_k = \mathrm{purchase} \\
  \land\, t_k \in [t_{\mathrm{now}} - \Delta T, t_{\mathrm{now}}] \big\}.
\end{split}
\end{equation}
\end{definition}

It is crucial to distinguish dormant users from cold-start users or low-interaction users commonly discussed in prior research.
While cold-start users typically lack interaction history (i.e., $|S_u| \approx 0$), dormant users on e-commerce platforms exhibit moderate browsing behavior.
However, they suffer from severe \textit{data sparsity}, with average click sequence lengths being approximately 14\% of that in the general population, which hinders prediction model training and further creates a \textit{self-reinforcing feedback loop}.

To bridge the gap between mere browsing and eventual conversion, we introduce the concept of \textit{functional roles} to model the latent trajectory of user intent.

%【lisk】：首先我觉得这里讲低U的问题时，可以简单写，《It is crucial to distinguish》这段有点长。以及我觉得你说这些是想引出functional role。这样的话，我觉得写《$|S_u| \approx 0$》不太合适。这里反而可以指出，低U用户有一些点击行为，是购买行为少。
%【lisk】：其实《To bridge the gap between mere browsing and eventual conversion,》这句已经有种“低u点击行为多、购买行为少”的意思了，但是这里可以写的写的更直白。毕竟functional role主要解决的问题就是，购买对于点击、曝光来说过于稀疏，以及购买行为的时间窗口很长。我们想用functional role，实时地归因每次点击，对于低u最终成交的价值。【这也是functional role的核心思想，介绍functional role的时候可以多往这个方向写】

\begin{definition}[Functional Role Trajectory]
Given the interaction sequence $S_u$ of user $u$, we extract a subsequence of key items $I^{\mathrm{key}}_u = [\tilde{i}_1, \tilde{i}_2, \dots, \tilde{i}_{M_u}]$ that drive the user's decision-making process. These key items are then abstracted into a \textit{functional role trajectory} 
\begin{footnotesize}
\begin{equation}
R_u = [\bm{r}_u(\tilde{i}_m)]_{m=1}^{M_u},
\end{equation}
\end{footnotesize}
where each $\bm{r}_u(\tilde{i}_m) \in \mathcal{R}$ represents the context-dependent utility (e.g., complement, substitute) of the corresponding item along the user's conversion path.
\end{definition}

% Furthermore, this creates a self-reinforcing feedback loop: suboptimal recommendations fail to elicit purchase signals needed for model correction, causing the system to collapse into safe but irrelevant popularity-biased recommendations.

% To break this cycle, our goal is to develop a reasoning-augmented generative framework that infers functional role trajectories $R_u$, enabling the system to capture latent user intent grounded in collective behavioral patterns, thereby effectively facilitating the transition of dormant users in $\mathcal{U}_0$ into active, converting users.

% To break this cycle, our primary objective is to infer the latent functional role trajectory $R_u$ grounded in collective behavioral patterns.
% By modeling $P(R_u | S_u)$, we aim to bridge the gap between sparse interactions and purchase intent, thereby facilitating the transition of dormant users into active buyers.
% \begin{definition}[Functional Role Trajectory]
% To capture the underlying intent behind item interactions, we introduce a set of latent functional roles $\mathcal{R}$ (e.g., \mathrm{trial, core, upgrade}).
% For each user $u$, the observed item sequence $S_u$ is governed by a latent \textit{Functional Role Trajectory} $R_u = [r_1, r_2, \dots, r_n]$, where each $r_k \in \mathcal{R}$ represents the semantic utility of the corresponding item $i_k$.
% \end{definition}

\begin{figure*}[!t]
    \centering
    \vspace{-10pt}
    \includegraphics[width=0.8\textwidth, trim=-0 0 -0 0, clip]{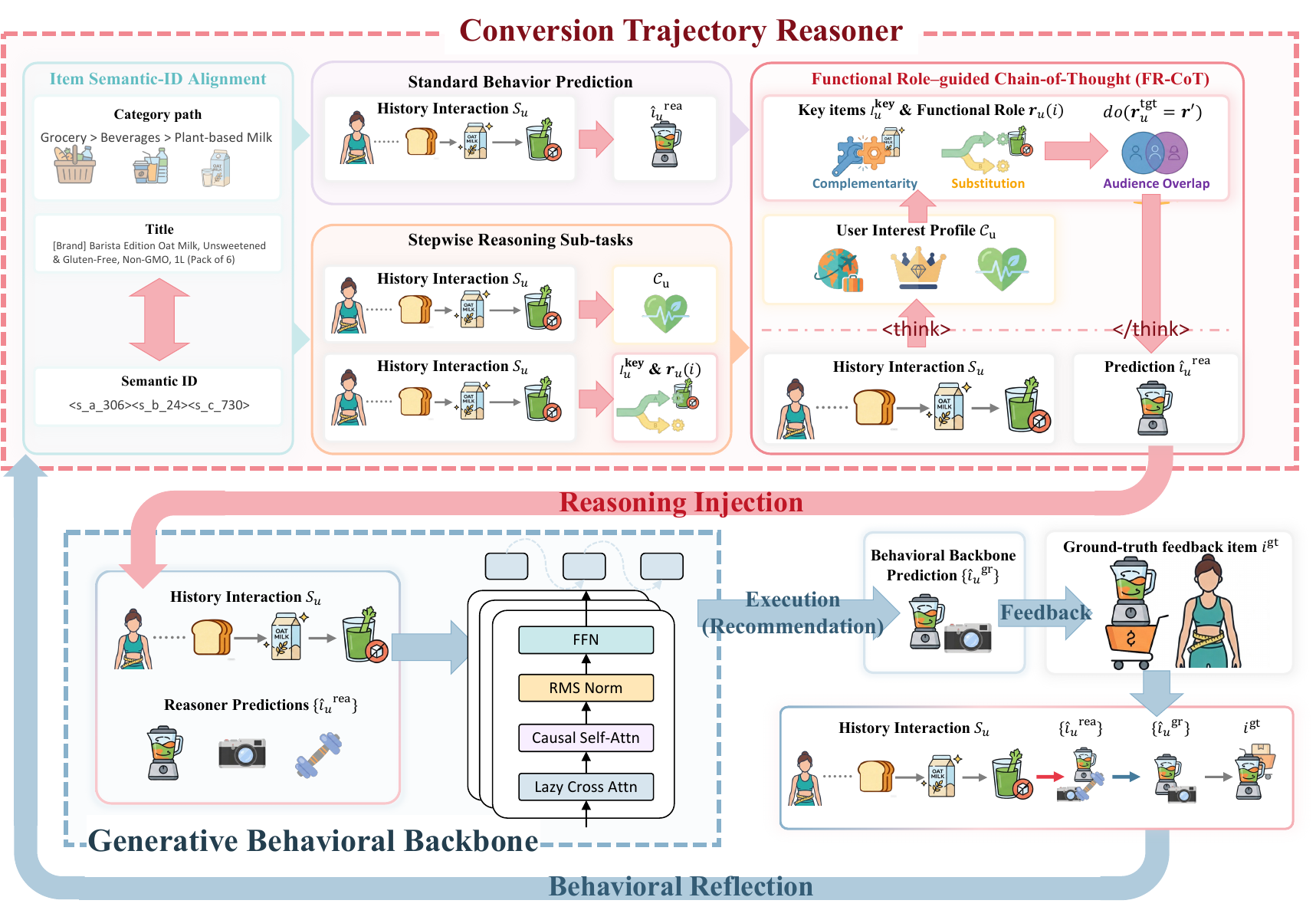}
    \vspace{-10pt}
    \caption{The overall architecture of \shortname{}. The framework consists of two synergistic modules: (1) The \textbf{Conversion Trajectory Reasoner} (top), which infers the latent \textit{Functional Role Trajectory} from sparse user behaviors and employs Counterfactual Inference to explore diverse conversion paths; and (2) The \textbf{Generative Behavioral Executor} (bottom), which grounds these reasoning signals into precise item retrieval via a generative backbone. The two modules are optimized through a collaborative ``reasoning–execution–feedback–reflection'' closed-loop training strategy.}
    \vspace{-5pt}
    \label{fig:model}
\end{figure*}

\subsection{Overview}
% objective：学习以往的u0用户是如何跃迁的，使得P(purchase | 序列，path)最大化；双轮驱动
% 为了打破解决数据稀疏问题、打破feedback loop，有必要引入更多额外知识来捕获这些u0用户的真实意图。对此我们提出\shortname{},如图\ref{fig:model}所示，其包含两个互相促进【todo：这个词语待定，从双轮驱动、互相帮助的角度找一个更精准有力量的】的模块，通过基于 collective behavioral patterns推理用户的latent user intent，以促进u0用户的转化。
% （1）世界知识大模型（大语言模型等backbone）【todo：这个名字会不会太普通了？要不要起一个易于记忆、又贴合论文创新点的名字】：这个模块的目标是利用世界知识和推理能力来挖掘u0用户发生转化的过程中的内在底层逻辑。LLM将在多种对齐和微调的SFT任务上训练，并通过反事实推理来推断符合用户跃迁路径的下一个可能会购买物品(s)。
% （2）行为大模型（生成式等推荐模型）：以平台上的协同行为数据进行训练，输入是用户的交互序列、用户、物品的side info、上下文信息等，输出是NTP。其中输入的交互序列中augment LLM生成的物品，以缓解数据稀疏问题和作为用户意图的显式补充。行为大模型的推荐商品及其对应的真实交互行为，也会输入LLM完成re-think。
% 综合以上，LLM推断的最符合跃迁路径的下一个item 增广至行为大模型，行为大模型的推荐及其反馈结果用于LLM反思，由此形成双轮驱动【todo：这个词语太像互联网公司用词，而非是学术用语。学术论文这类互相迭代的模式一般有什么好的命名建议】。需要注意的是，行为大模型（即纯CF-based 推荐模型）在这里是不可或缺的。因为推荐本质上仍是一个基于CF而非商品本身知识的系统，仅使用LLM进行NTP的话，此时模型中依然是语义知识或者说世界知识占主导，其优势是泛化性而非准确性。【todo：这样讲会不会不太好？显得大模型不好（但不够纯粹准确又是事实）】

% virtuous closed-loop mechanism用来对应之前挑战里说传统方法跳不出loop

% todo: 检查一下逻辑承接是否通顺（有没有一些术语从来没提过就出现、跟之前problem或自己的上下文有没有冗余表述）

% To address the data sparsity issue and break the feedback loop in dormant user scenarios, it is essential to incorporate external knowledge to uncover latent user intent. To this end, we propose \shortname{}, a unified framework illustrated in Figure \ref{fig:model}. The framework consists of two synergistically coupled modules that work in tandem to infer latent intent grounded in collective behavioral patterns, aiming to facilitate the conversion transition of dormant users. 

To achieve this objective, we propose \shortname{}, a unified reasoning-augmented generative framework illustrated in Figure \ref{fig:model}.
The framework consists of two synergistically coupled modules that work in tandem to uncover latent user intent, utilizing external knowledge to drive the conversion of dormant users.

\textls[-18]{\textbf{(1) Conversion Trajectory Reasoner.}
Acting as an intent planner, this module leverages the semantic reasoning capabilities of LLMs to distill the underlying decision logic behind dormant-to-active transitions. 
Trained on multi-task supervised fine-tuning (SFT) objectives, the Reasoner performs counterfactual reasoning to infer plausible functional role trajectories derived from successful precedents, generating potential next-items that align with these transition paths.}

\textls[-12]{\textbf{(2) Generative Behavioral Backbone.}
This module functions as the execution unit, formulated as a generative collaborative filtering model.
Despite the semantic power of LLMs, reliable recommendation in industrial settings fundamentally relies on platform-specific collaborative signals. 
Therefore, the backbone complements the Reasoner by grounding semantic predictions in large-scale interaction data.
Crucially, the two modules establish a virtuous closed-loop. The interaction sequences fed into this backbone are augmented with the items generated by the Reasoner, serving as an explicit supplement to user intent to alleviate sparsity.
In turn, the backbone's recommendations and the subsequent collected user responses are fed back into the Reasoner to trigger a \emph{reflection} mechanism.}

% While both modules predict the next item, the behavioral backbone (i.e., the CF-based recommender) remains indispensable in this architecture.
% Fundamentally, recommendation relies heavily on \textit{collaborative-filtering signals} (user–item interaction patterns) rather than purely on \textit{item semantics}.
% When used alone for next-item prediction, LLMs tend to overemphasize semantic and world knowledge while under-utilizing platform-specific collaborative signals.
% The behavioral model complements the LLM by grounding predictions in large-scale interaction data, ensuring that recommended items are not only semantically plausible but also consistent with collaborative pattern.

\vspace{0.5em}
\noindent\textbf{Reasoner Outline.}
\textls[-12]{To empower the Reasoner with the conversion trajectory reasoning capabilities, we adopt a multi-stage instruction-tuning paradigm that progressively internalizes platform-specific knowledge~\cite{ye2025align3gr,liu2025onerecthink,zeng2025grm}:
% we adopt a multi-stage instruction-tuning paradigm inspired by recent work~\cite{ye2025align3gr,liu2025onerecthink,yi2025recgpt,zeng2025grm}. 
% This paradigm progressively internalizes platform-specific world knowledge and collective behavioral patterns, enabling the inference of latent interests from sparse and noisy behavior sequences.
% The subsequent sections detail this process:
(1) \textbf{item understanding}, bridging the gap between general linguistic knowledge and platform-specific identifiers to establish a cognitive foundation (Section \ref{sec:item_align});
(2) \textbf{functional-role-guided reasoning}, incorporating functional roles as structured reasoning intermediates to distill user intent trajectories for target item prediction (Section \ref{sec:func_role}); and
(3) \textbf{counterfactual inference}, applying treatments to target functional roles to diversify generation and mitigate interest collapse for dormant users (Section \ref{sec:cf}).}

%【lisk】：感觉《Crucially, it establishes a virtuous》和《While both modules predict the next item》这两段可以合并一下，都在讲我们要双轮驱动，协同信息为主，世界知识为辅。可以合并下，节约空间。以及这里既然讲协同信息为主，前面就更不该写《低U的$|S_u| \approx 0$》

% \subsection{Conversion Trajectory Reasoner}  
% \label{sec:reasoner}
% 【总体大纲：各种对齐任务；Functional role的抽象与path；Counterfactual Role Reasoning】

% 本节介绍如何微调LLM，用来学习用户共识，即平台-specific世界知识，来从稀少且噪声大的u0行为序列中推理u0兴趣，作为后续Generative backbone的输入补充。Following 【cite】等工作，我们的Reasoner将通过多阶段微调包含从语义理解、到行为建模、到反事实推理的三层能力，具体为：（1）Item Understanding ，（2）Functional Role Reasoning，（3）Counterfactual Inference Strategy。

% In this section, we describe how we fine-tune a general LLM to internalize platform-specific world knowledge and collective user behavioral patterns, thereby enabling it to infer latent user interests from sparse and noisy behavior sequences.
% Inspired by recent work~\cite{ye2025align3gr,liu2025onerecthink,yi2025recgpt,zeng2025grm}, we adopt a multi-stage instruction-tuning paradigm that progressively equips the LLM with three essential capabilities:

\subsection{Item Semantic-ID Alignment} 
\label{sec:item_align}

% 为了能让大模型预测next Item，首先需要大模型先认识Item，即将大模型原本的语义空间与平台的item identifier（ID）对齐，这是Prerequisite。这里整体包含Expanded Vocabulary和Multi-task SFT两步。（1）首先。我们使用Item的semantic ID （SID），并expand the LLM vocabulary to include the item SID tokens。这里的SID采用Kuaishou平台内部对齐的电商item SID~\cite{luo2025qarm_sid}，利用商品title、image等多模态信息以及协同过滤信息得到3级SID。（2）接着， 为了construct 对齐SFT的prompt，我们收集如下item meta信息：item Title，item price，item category（三级，例如“数码>手机及其配件>手机支架”）。在扩充SID的LLM上，我们设计以下multi-task Itemic Alignment prompt：item Prediction based on item information (item meta -> SID), item title prediction based on item SID (SID -> item title), item category prediction based on item SID (SID -> item category). To prevent catastrophic forgetting of general linguistic capabilities, we mix general instruction data (e.g., FLAN) with item alignment data

% 这个阶段的目标是让SID的语义空间与原语义空间对齐，因此训练时只训练LLM的token embedding和输出的lm head部分，其他部分冻结。

A prerequisite for LLM-based next-item recommendation is to enable the model to recognize items, i.e., to align its general semantic space with our platform's discrete item ID space. We accomplish this through two steps: vocabulary expansion and multi-task SFT.

First, we expand the LLM vocabulary to include platform-specific item tokens. 
Here, we adopt the hierarchical Semantic ID (SID) codebook deployed in Kuaishou's production environment~\cite{luo2025qarm_sid} as the unique item representation. 
These SIDs are constructed by encoding multi-modal features (e.g., titles, images) and collaborative signals, and then applying RQ-KMeans to obtain a three-level hierarchical code.
Second, to establish a bidirectional mapping between item semantics and SIDs, we construct a multi-task instruction-tuning dataset using rich item metadata, including titles, prices, and hierarchical category paths (e.g., \texttt{Electronics > Mobile Accessories > Phone Stand}).
% (e.g., \texttt{Electronics} > \texttt{Mobile Accessories} > \texttt{Phone Stand}).
The alignment tasks include:
\begin{itemize}[leftmargin=*, nosep]
    \item \hypertarget{para-item-indexing}{\textbf{Item Indexing}} (item metadata $\to$ SID), where the model predicts the corresponding SID given item attributes; and
    \item \hypertarget{para-item-profiling}{\textbf{Item Profiling}} (SID $\to$ item title, SID $\to$ item category), where the model reconstructs the title or category path given its SID.
\end{itemize}
To prevent catastrophic forgetting of general linguistic capabilities, we mix general instruction data~\cite{peng2023gpt4data, luo2025empirical} with these item alignment samples during training.

%【lisk】：从节约篇幅的角度，\subsection{Item Semantic-ID Alignment} 是否不需要单写一个章节？是否可以在Overview中的，\textbf{Reasoner Outline}之前，单起一段简要介绍\textbf{SID}，然后把item_indexing和item_profiling两个任务，放到Reasoner Outline 的 item understanding 部分？感觉就是3.2和3.3可以合并。感觉3.3的内容都不太重要。

% \subsection{Functional-Role-Guided Reasoning}
\subsection{Functional Role Modeling}
\label{sec:func_role}

\subsubsection{Motivation.} % 需要reasoning -> 需要structured reasoning -> 用fr -> 为什么选择fr
Following the alignment of item semantics, the next crucial step is to align the LLM with user–item interaction patterns~\cite{ye2025align3gr}. 
However, directly instructing the LLM to map interaction sequences to a target item is challenging for dormant users, as their behaviors are sparse, noisy, and lack explicit interest signals. %【lisk】：这里我还是觉得应该强调低U用户的转化链条很长，不够实时。有过很多点击后，得过段时间才可能有转化。导致了转化的归因耗时久。所以要FR。当然了，sparse、lack explicit signals这些都可以保留。
To bridge this gap, we leverage the LLM’s reasoning to distill user interests under structured guidance~\cite{zeng2025grm}.

Specifically, we introduce the \textit{Functional Role Trajectory} as an abstraction of user intent evolution. We posit that user interactions intrinsically follow regular decision patterns that can be abstracted and transferred across users and items. For example, a click sequence like "Breakfast Milk $\to$ Coffee $\to$ Coffee Machine" reflects an underlying logic of "Daily Necessity $\to$ Scenario Complement $\to$ Extension into High-Value Items". Modeling such trajectories as a Chain-of-Thought offers two key advantages: (1) Intuitive Interpretability: It aligns with how humans naturally reason about purchasing decisions. While items like "Milk" and "Kitchen Paper"  differ visually, they share the same functional role as "Daily Necessities" in a user's decision path. (2) Statistical Efficiency: Compared to the sparse and large-cardinality item space, the functional role trajectory operates in a compact space with significantly higher pattern recurrence, making it easier for the model to capture transferable behavioral modes and generalize from limited signals.

\subsubsection{Functional Role Definition}
\textls[-12]{In user $u$'s sequence, we denote an item $i$'s functional role $\bm{r}_u(i) \in \mathcal{R}$ as a multi-dimensional} discrete vector
\begin{footnotesize}
\begin{equation}
\label{eq:role}
\bm{r}_u(i) = \left( r_{\mathrm{pop}}(i),\, r_{\mathrm{cost}}(i),\, r_{\mathrm{repl}}(i),\, r_{\mathrm{rel}}(i; \mathcal{C}_u) \right),
\end{equation}
\end{footnotesize}
which concatenates intrinsic item attributes and context-dependent relational roles. We summarize the construction below.

\textbf{Intrinsic Attribute Roles.}
These roles characterize the inherent static properties of an item. We distill them into three dimensions:
\begin{itemize}[leftmargin=*, nosep]
  \item \textbf{Market Popularity} $r_{\mathrm{pop}}(i)$: reflects the item-level collective user preference. We discretize popularity into three levels based on category-normalized item sales: 
  \begin{itemize}[leftmargin=1.5em]
  \item \textit{Booming} (high-velocity trends), 
  \item \textit{Evergreen} (consistent sellers), and 
  \item \textit{Long-tail} (niche demands).
  \end{itemize}
  \item \textbf{Replenishment Nature} $r_{\mathrm{repl}}(i)$: captures the category-level recurrence frequency of the underlying user need. Based on the aggregate sales volume of the item's category, we categorize items into:
  \begin{itemize}[leftmargin=1.5em]
  \item \textit{FMCG} (Fast-Moving Consumer Goods, e.g., tissue) versus 
  \item \textit{Durables} (e.g., appliances).
  \end{itemize}
  \item \textbf{Decision Cost} $r_{\mathrm{cost}}(i)$: reflects the cognitive effort and perceived risk associated with the purchase. Considering both the raw price and its category price distribution, we classify items into:
  \begin{itemize}[leftmargin=1.5em]
  \item \textit{Trial}: low-cost items that facilitate trust-building and impulsive conversion;
  \item \textit{Core}: standard consumption choices; and
  \item \textit{Premium}: high-price items that typically require repeated comparison and deliberation (e.g., expensive electronics).
  \end{itemize}
\end{itemize}

% c_u解释 todo
\textbf{Contextual Intent Role.}
\textls[-12]{Intrinsic roles alone cannot capture how an item relates to the user’s dynamic intent in the current sequence $S_u$. Thus, we introduce a contextual relational role $r_{\mathrm{rel}}(i; \mathcal{C}_u)$ defined with respect to the user's cumulative interest profile $\mathcal{C}_u$. 
First, we aggregate behavior-weighted scores over historical categories in the interaction sequence $S_u$, and obtain an ordered interest profile $\mathcal{C}_u = [c_k]_{k=1}^{L_u}$, where $c_k$ denotes the $k$-th category in user $u$'s behavioral interest ranking and $L_u = |\mathcal{C}_u|$.
Second, we construct a domain-specific e-commerce decision knowledge graph from large-scale interaction logs on our platform, encoding semantic and behavioral relations between categories. 
We define three relation types: }
% \textit{Complementarity} (scenario completion), \textit{Substitution} (functional alternatives), and \textit{Audience Co-reference} (similar user demographics and cohort preferences). 
\begin{itemize}[leftmargin=2.5em,label=--]
\item \textit{Complementarity} (scenario completion), 
\item \textit{Substitution} (functional alternatives), and 
\item \textit{Audience Overlap} (appealing to similar user personas). 
\end{itemize}
For example, the edge “Phone $\to$ Phone Stand” is labeled as a \textit{Complementarity} relation. 
Finally, the item's \textbf{Contextual Intent Role} $r_{\mathrm{rel}}(i; \mathcal{C}_u)$ is determined by the dominant relation between $c(i)$ and the high-scoring categories in the user’s interest profile $\mathcal{C}_u$ via this graph:
\begin{equation}
\label{eq:rel_role}
\begin{aligned}
    r_{\mathrm{rel}}(i; \mathcal{C}_u) &  = \left(c^*, \mathrm{Rel}_{c^* \to c(i)} \right), \\  
    c^* & = \mathop{\arg\max}_{c \in \mathcal{C}_u}\bigl(\mathrm{Score}_u(c) \cdot \mathbb{I}[\,\exists \mathrm{Rel}_{c \to c(i)}]\bigr),
\end{aligned}
\end{equation}
where $c(i)$ denotes the category of item $i$, $\mathrm{Rel}_{c^* \to c(i)}$ represents the relation type on the edge, and $\mathrm{Score}_u(c)$ denotes the behavior-weighted interest score of user $u$ for category $c$.

%【lisk】：关于《\textbf{Intrinsic Attribute Roles.}》 和 《\textbf{Contextual Intent Role.}》这两段：我感觉这两段的介绍顺序有较大的问题。如果我来介绍的话，1. 我会先讲一下c2c关联，表示是用deepseek满血版构造的，有几种关系，后续可以说我们在蒸馏deepseek。 2. 给定用户历史序列，找到user的优势类目，c2c扩展到核心item，列出核心item与优势类目的c2c关联。3. 核心item的 intrinsic attributes。另外这里也需要精简下，不行就把一些东西放附录？

\subsubsection{Role Trajectory Construction.}
With the multi-dimensional role $\mathbf{r}_u(i)$ fully specified (Eq.~\eqref{eq:role}), we proceed to construct the reasoning trajectory.
For each user sequence $S_u$, we filter interactions to retain a subsequence of key items $I^{\mathrm{key}}_u = [\tilde{i}_1, \tilde{i}_2, \dots, \tilde{i}_{M_u}]$, each of which possesses a valid contextual role as defined in Eq.~\eqref{eq:rel_role}.
By mapping each key item $\tilde{i}_m$ to its instantiated role vector, we obtain the final \textbf{Functional Role Trajectory}:
\begin{equation}
\label{eq:role_traj}
    R_u = \left[ \bm{r}_u(\tilde{i}_1), \bm{r}_u(\tilde{i}_2), \dots, \bm{r}_u(\tilde{i}_{M_u}) \right].
\end{equation}
This explicit trajectory summarizes how the user's needs and decision context evolve along the conversion path: for example, items with $r_{\mathrm{repl}}(i)=\text{FMCG}$ and $r_{\mathrm{cost}}(i)=\text{Trial}$ typically correspond to a “Daily Necessity” intent.

\vspace{0.5em}

Crucially, our work differs fundamentally from prior CoT approaches in recommendation~\cite{yi2025recgpt, liu2025onerecthink, tang2025lrem}. Existing methods typically summarize \textit{unilateral} information (e.g., disjoint item tags or static user profiles). In contrast, our Functional Role captures \textit{bilateral interaction information}, a cross-feature synthesis of item characteristics and user interests.
This bilateral coupling provides a more faithful and transferable signal, enabling the LLM to reason about \textit{why} a user interacts, rather than merely \textit{what} they interact with.

\subsection{Instruction Tuning with Structured Reasoning}
Based on the functional role trajectory $R_u$, we design a \textbf{Functional Role–guided Chain-of-Thought} (FR-CoT) objective that formulates dormant-user conversion as a structured reasoning problem, as illustrated in Fig.~\ref{fig:model}.

For each user $u$, the FR-CoT input consists of the user profile and the SID-indexed interaction sequence $S_u$, while the target is a chain-of-thought (CoT) style output wrapped in \texttt{<think>} tags, followed by the target item SID. 
The CoT is instructed to follow a three-step reasoning logic:
\begin{enumerate}[label=(\arabic*), leftmargin=*]
    \item distilling the user's current interest profile $\mathcal{C}_u$;
    \item identifying key items $I^{\mathrm{key}}_u$ and their functional roles to reconstruct the role trajectory $R_u$; and
    \item explicitly inferring the functional role of the target item, denoted as $\bm{\hat{r}}^{\text{tgt}}_u$.
\end{enumerate}
Crucially, predicting $\bm{\hat{r}}^{\text{tgt}}_u$ before the item ID serves as an intent anchor, which effectively prunes the search space and constrains generation to items consistent with the inferred conversion logic (e.g., encouraging transitions such as \textit{Trial} $\to$ \textit{Premium} when supported by the trajectory).
%【lisk】：crucially这段，是不是可以说，FR对于user的主要兴趣做了个summary，辅助LLM在主要兴趣上做推理？从这个角度上说，前面的低U面临的问题中，确实要提一下sparsity问题。

To enhance reasoning robustness, we optimize the FR-CoT objective jointly with a set of auxiliary SFT tasks.
For notational simplicity, we denote the input context (comprising both the user profile and interaction history) simply as $S_u$ below.
The auxiliary tasks are categorized into three objectives:
\begin{itemize}[leftmargin=*, nosep]
    \item \hypertarget{para-standard-behavior-prediction}{\textbf{Standard behavior prediction.}} We retain standard sequence-to-item tasks to train the model with direct collaborative filtering capabilities, 
    including $S_u \to$ target SID and $S_u \to$ target title. 
    
    \item \hypertarget{para-stepwise-reasoning-sub-tasks}{\textbf{Stepwise reasoning sub-tasks.} }
    To enforce the quality of intermediate reasoning, we decompose the full CoT into independent sub-tasks, including 
    $S_u \to I^{\mathrm{key}}_u$ to localize key items, 
    $(S_u, I^{\mathrm{key}}_u) \to R_u$ to reconstruct the functional role trajectory and, 
    $(S_u, I^{\mathrm{key}}_u, R_u, \bm{r}^{\text{tgt}}) \to$ Target SID to ground prediction on the target role.  
    These objectives teach the model to separately internalize intermediate concepts before composing them in full FR-CoT reasoning.
    
    \item \hypertarget{para-semantic-alignment-replay}{\textbf{Semantic alignment replay.} }
    We incorporate \textit{Item Indexing} (Item metadata $\to$ SID) and \textit{Item Profiling} (SID $\to$ item title, SID $\to$ item category) tasks from Section~\ref{sec:item_align} to prevent catastrophic forgetting of the index–language alignment.
\end{itemize}
All tasks are optimized jointly with the main FR-CoT objective, and detailed prompt templates are provided in Appendix~\ref{app:prompt}.

% 【lisk】：Semantic alignment replay 是不是应该写在three objectives中的第一位？

\subsection{Counterfactual Functional Role Inference}
\label{sec:cf}
% 为了对后续Generative backbone进行增广的时候能够尽可能多增加数据，在LLM inference阶段，我们需要对结果进行beam search来获得同一个user的多个可能target item。然而过长的reasoning文本使得采样和beam search的推理成本很高；而只在最终生成target item的时候才beam search又会多样性下降，因为已经有target Functional role的约束。为此，我们的策略是对target Functional role进行treatment，其反事实的值设置为当前seq中以及常见target functional role中的随机采样值。with 这个counterfactual值，我们使用seq + reasoning to next item任务进行推理，如图\ref{fig:role}所示。
% Counterfactual Inference的优势是以极低的修改成本，配合着reasoning部分的Functional role trajectory的存在，在符合用户转化逻辑的情况下增加推理结果的多样性。这对u0用户是尤其重要的，因为他们的交互通常噪声很大，简单重复过往兴趣进行推荐可能会陷入信息茧房，也就是推荐系统困在不符合用户真实意图的self-reinforcement feedback loop中一直无法转化；而反事实的Functional role inference增加了跳出这个loop的可能性。

% {Motivation} 即使我们利用LLM的reasoning能力进行推理，其推断的target Functional role依然容易陷入Self-reinforcing loop。这是因为沉睡用户本身交互稀疏且噪声大，依赖maximum likelihood estimation (MLE)的训练推理会使得LLM overfit monotonous历史序列中的主导兴趣，导致推荐单一，难以predict真正的兴趣而唤醒。

% 尽管我们可以使用Beam search来增加结果的多样性，但是面临以下两个问题：（1）\textit{Computational Cost}: beam search over the entire Chain-of-Thought sequence is prohibitively expensive due to the length of reasoning traces， 也难以约束结果的合法性; （2） \textit{Diversity Collapse}: Performing beam search only at the final item generation step is ineffective, as the generation is already strongly constrained by the predicted target role $\bm{\hat{r}}^{\text{tgt}}_u$ (the intent anchor), resulting in homogenous outputs.

\subsubsection{Motivation.}
Even with FR-CoT reasoning, the predicted target functional role for a dormant user may still be trapped in a self-reinforcing loop.
Given that dormant users exhibit sparse and noisy interactions, maximum likelihood estimation (MLE) tends to overfit the dominant but possibly stale interests in $S_u$, leading to interest collapse and reinforcing the Matthew effect.
Standard decoding strategies like Beam Search offer limited remedy, as they operate at the token level. Once the high-level intent anchor (predicted target role $\hat{\bm{r}}^{\mathrm{tgt}}_u$) is determined by the Reasoner, the diversity of subsequent item generation is structurally bounded (see Section ~\ref{sec:case}).
To break this loop, we argue that exploration must occur at the \textit{intent level} rather than the \textit{item level}.

% A straightforward remedy is to apply beam search. However, this faces two challenges:
% (1) \textit{Computational and structural constraints}: beam search over the entire CoT sequence is prohibitively expensive due to long reasoning traces, and it is difficult to ensure that the generated CoT remains well-formed and respects the required structured reasoning format;
% (2) \textit{Diversity collapse}: applying beam search only at the final item token yields limited diversity, since generation is already strongly constrained by the predicted target role $\hat{\bm{r}}^{\mathrm{tgt}}_u$ (the intent anchor), leading to highly similar candidate items.

\subsubsection{Solution.}
To address these issues, we propose \textbf{Counterfactual Functional Role Inference}, which diversifies recommendations by intervening on the intent anchor in the compact functional role space rather than on the full token sequence.
Instead of strictly conditioning on the FR-CoT prediction $\hat{\bm{r}}^{\mathrm{tgt}}_u$, we apply a counterfactual intervention $do(\bm{r}^{\mathrm{tgt}}_u = \bm{r}')$. The counterfactual role $\bm{r}'$ is sampled from a candidate set of target roles $\widetilde{\mathcal{R}}^{\mathrm{tgt}}_u$ constructed by combining:
(i) roles already observed along the role trajectory $R_u$; and
(ii) globally frequent conversion roles mined from historically successful transitions.
We then reuse the FR-CoT prompt while substituting the target role with the counterfactual one $\bm{r}'$, and let the Reasoner predict the corresponding next item $\hat{i}^{\mathrm{rea}}_u$:
\begin{footnotesize}
\begin{equation}
\label{eq:cf} 
    \underbrace{(u, S_u)}_{\text{Context}}, \;
    \texttt{<think>} 
    \underbrace{\mathcal{C}_u}_{\text{Intent}}, \;
    \underbrace{R_u}_{\text{Trajectory}}, \;
    \underbrace{\bm{r}' \sim P(\widetilde{\mathcal{R}}^{\mathrm{tgt}}_u)}_{\text{CF Intervention}}, \;
    \texttt{</think>} \;\xrightarrow{\quad\mathbf{Reasoner}\quad} \;
    \hat{i}^{\mathrm{rea}}_u.
\end{equation}
\end{footnotesize}
Conceptually, this process instructs the LLM: \textit{“If the user were to act under target role $\bm{r}'$, which item would be most likely to lead to conversion?”}
Subsequently, we employ beam search decoding to generate a diverse set of candidate items $\{\hat{i}^{\mathrm{rea}}_u\}$ based on this counterfactual premise.

This counterfactual mechanism provides two key benefits:
\begin{itemize}[leftmargin=*, nosep] 
    \item Logic-constrained exploration.
    Intervening on the target role yields \emph{structured diversity}. The model explores alternative conversion paths (e.g., \textit{Trial} $\to$ \textit{Premium} upgrade vs.\ \textit{Trial} $\to$ \textit{FMCG} replenishment), while remaining consistent with the user’s inferred intent and the upstream FR-CoT reasoning.
    \item Efficient multi-candidate generation.
    Since the intervention operates in a low-cardinality role space rather than recomputing entire reasoning traces, multiple counterfactual candidates can be generated in parallel with minimal additional cost, effectively breaking repetitive self-reinforcing loops for dormant users and offering a cost-effective exploration–exploitation trade-off.
\end{itemize}

\subsection{Generative Behavioral Backbone}
\label{sec:gen_struct}

%【lisk】：这段没啥创新，因此还可以再删减一点。
% 介绍我的生成式推荐框架
% While the Reasoner can directly output next-item candidates, relying solely on it for final ranking entails risks in industrial settings. LLMs predictions are prone to semantic bias and may drift away from implicit collaborative-filtering (CF) patterns.
% To bridge the gap between semantic reasoning and behavioral accuracy, we adopt a dual-module collaborative design: the Reasoner serves as a knowledge-guided planner, and a CF-based recommendation backbone acts as the execution engine.

Although the Reasoner captures user intent, relying solely on it for online serving is suboptimal. 
LLM predictions often drift from implicit collaborative-filtering (CF) patterns, and their prohibitive inference latency hinders real-time response, while caching strategies inevitably introduce data staleness.
To address these limitations, we employ a Generative Behavioral Backbone as the execution engine, utilizing the Reasoner as a knowledge-guided planner.

\textls[-22]{Specifically, we instantiate the backbone using a state-of-the-art decoder-only generative architecture~\cite{zhou2025onerecv2}. By representing items as hierarchical SIDs, this approach transforms recommendation into a structured token generation task. This formulation significantly lowers the learning difficulty compared to traditional flat softmax over the full item space and facilitates the seamless injection of the Reasoner's guidance.}

In the behavior backbone, we concatenate multiple user-side features (e.g., static profile, interaction history) and linearly map them into a unified \textit{Context} representation $\bm{T}$.
To ensure efficiency, $\bm{T}$ is normalized via RMSNorm to yield static key-value pairs $(\bm{k}, \bm{v})$, which serve as a persistent memory cache accessible to the decoder throughout the entire generation process.
The decoder consists of $L_\mathrm{decoder}$ stacked transformer blocks that generate the target item's SIDs autoregressively.
Crucially, to optimize inference, each block employs a lazy cross-attention mechanism~\cite{zhou2025onerecv2} that attends to the shared static cache $(\bm{k}, \bm{v})$ without redundant projections, followed by standard causal self-attention and position-wise FFNs.

% The decoder comprises $L_\mathrm{decoder}$ stacked transformer blocks that generate the target item's SIDs autoregressively. 
% Crucially, each decoder block employs a lazy cross-attention mechanism that directly attends to the cached static representations from the encoding module without redundant projection. 
% The hidden state $\bm{h}^{(l)}$ at layer $j$ is updated as follows:
% \begin{footnotesize}
% \begin{equation}
% \begin{aligned}
%     \bm{h}_{\text{cross}}^{(l)} &= \bm{h}^{(j-1)} + \mathrm{CrossAttn}\Bigl(\mathrm{RMSNorm}\bigl(\bm{h}^{(j-1)}\bigr), \bm{k}_{\phi(l)}, \bm{v}_{\phi(l)}\Bigr), \\
%     \bm{h}_{\text{self}}^{(l)}  &= \bm{h}_{\text{cross}}^{(l)} + \mathrm{SelfAttn}\Bigl(\mathrm{RMSNorm}\bigl(\bm{h}_{\text{cross}}^{(l)}\bigr)\Bigr), \\
%     \bm{h}^{(l)}                &= \bm{h}_{\text{self}}^{(l)} + \mathrm{FFN}^{(l)}\Bigl(\mathrm{RMSNorm}\bigl(\bm{h}_{\text{self}}^{(l)}\bigr)\Bigr),
% \end{aligned}
% \label{eq:decoder_block}
% \end{equation}
% \end{footnotesize}
% where $\phi(l)$ maps the current decoder layer to the corresponding shared context KV pair. 

%【lisk】：这个公式里的crossattn应该是causual-attn？你确认下。就是mask前面的元素。

% Without loss of generality, we employ OneRec~\cite{deng2025onerec} as the representative backbone in our implementation for its SOTA performance. Yet our framework is backbone-agnostic that compatible with other recommendation models (Section~\ref{sec:ex_backbone}).

\subsection{Co-Training Strategy: Reasoning Injection and Behavioral Reflection}  % todo:标题的大概意思是双轮驱动
\label{sec:synergy}
% Injection" from Reasoner to Backbone and the "Reflection" from Backbone back to Reasoner
To fully bridge the gap between semantic reasoning and collaborative execution, we establish a bidirectional closed-loop between the Reasoner and the Behavioral Backbone. 
This mechanism ensures that the Reasoner's high-level insights effectively guide the backbone's generation, while the backbone's interaction feedback continuously refines the Reasoner's logic.

\textbf{Forward Guidance via Reasoning Injection.}
\textls[-18]{A primary challenge in the behavioral module is the scarcity of reliable signals for dormant users. To alleviate this, we inject the Reasoner’s counterfactual reasoning into the generative backbone.
For user $u$, we encode the Reasoner's predictions $\{\hat{i}^{\mathrm{rea}}_u\}$ into structured \textbf{Reasoning Guidance Features}, including (i) the raw candidate item list; and (ii) hierarchical semantic statistics, such as the positional patterns of coarse-grained SIDs (e.g., the first two layers).
These features are concatenated into the backbone's input context, guiding the generation of the final recommendation set $\mathcal{I}^{\mathrm{gr}}_u$ with explicit semantic navigational cues.}

%【lisk】：这里是否要提一下，我们用的是Qwen生成的2级sid，及其统计信息呢？主要突出下《2级sid》？

\textbf{Backward Alignment via Behavioral Reflection.}
To further harmonize the collaboration between the semantic Reasoner and the behavioral backbone, we introduce a \textit{Reflection} mechanism that utilizes the backbone as a collaborative mirror to rectify the Reasoner.
We construct a post-hoc instruction-tuning task using collected online logs. 
Specifically, given the original user profile and behavior sequence $S_u$, the two modules sequentially produce their recommendations $\{\hat{i}^{\mathrm{rea}}_u\}$ and $\{\hat{i}^{\mathrm{gr}}_u\}$. From the online exposure of $\{\hat{i}^{\mathrm{gr}}_u\}$, we can observe the ground-truth feedback item $i^{\mathrm{gt}}$.
We then fine-tune the Reasoner to recover $i^{\mathrm{gt}}$ from this enriched context via the following instruction tuning task:
\begin{footnotesize}
\begin{equation}
\label{eq:reflection}
(S_u, \{\hat{i}^{\mathrm{rea}}_u\}, \{\hat{i}^{\mathrm{gr}}_u\}) \xrightarrow{\text{Reasoner Reflection}} i^{\mathrm{gt}}.
\end{equation}
\end{footnotesize}
\textls[-17]{Through this reflection process, the Reasoner iteratively calibrates its semantic logic against actual collaborative evidence, fostering a virtuous cycle that synchronizes reasoning with recommendation execution.}

\begin{figure}[!t]
    \centering
    \vspace{-12pt}
    \includegraphics[width=0.45\textwidth, trim=0 5 0 0, clip]{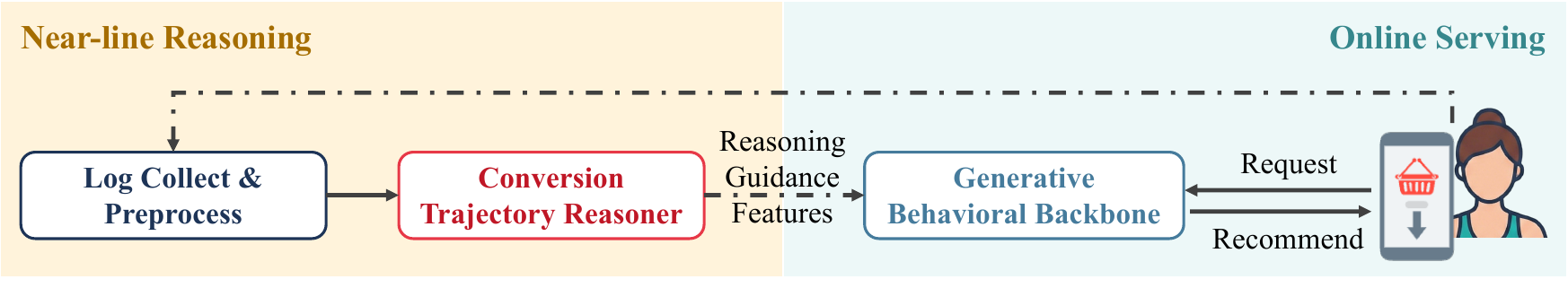}
    \vspace{-12pt}
    \caption{Framework of Online Deployment of \shortname{}. }
    \vspace{-16pt}
    \label{fig:deployment}
\end{figure}

\section{Implementation and System Deployment}

In this section, we outline the practical implementation of \shortname{} on the Kuaishou e-commerce platform.
Detailed training configurations, including optimization function and specific parameter freezing schedules, are provided in Appendix~\ref{app:implementation}. Here, we highlight a key Curriculum Learning strategy adopted during the item alignment phase to ensure robust semantic representation.

\noindent\textbf{Curriculum Alignment Strategy.}
Instead of one-shot training, we align items in a progressive manner:

\begin{itemize}[leftmargin=*, nosep]
\item {Global Alignment}: We first train on the full item set to establish a broad cognitive foundation, ensuring the model covers the entire distribution of item semantics;
\item {High-Quality Refinement}: We then fine-tune exclusively on a subset of high-quality items (filtered by interaction feedback).
\end{itemize}
Crucially, this strategy prevents noisy, low-quality items from being treated indiscriminately with high-quality ones, thereby minimizing noise propagation and enhancing both the accuracy and robustness of semantic alignment.

\vspace{0.5em}
\noindent\textbf{System Deployment.}
\label{sec:deployment}
\shortname{} has been deployed on the Kuaishou e-commerce platform, which serves over 700 million monthly active users (MAU). 
As illustrated in Figure~\ref{fig:deployment}, the deployment architecture follows an asynchronous collaborative design: the Reasoner is scheduled for weekly updates and inference, while the behavioral model supports online streaming updates and real-time serving.
We provide comprehensive details regarding the near-line reasoning pipeline and online serving configurations in Appendix~\ref{app:deploy_details}.

\section{Experiments}

In this section, we present comprehensive offline and online evaluations of \shortname{}, aiming to answer the following key questions:
\begin{itemize}[leftmargin=*, nosep]
\item \textbf{RQ1:} How does \shortname{} perform compared with strong recommendation baselines for dormant users? % base(例如u2i、i2i)【todo】、onerec【done】、qwen+onere【done】离线指标对比；Reasoner【todo】对比SASRec【todo】，Tiger【todo】的推理差异
\item \textbf{RQ2:} How do the key design components of \shortname{}, in particular the Reasoner and its Functional-Role-guided reasoning, contribute to the overall performance?  % qwen不同SFT阶段（ckpt）对于各个任务的指标差异；不同数据规模的SFT对结果指标的影响【todo】
\item \textbf{RQ3:} How effectively and robustly does the Reasoner guide the behavioral backbone to break the self-reinforcing feedback loop?  % 马太效应的分布对比onerec，qwen+onerec【done】; SASRec【todo】, tiger【todo】, qwen【done】对于OOD数据的结果差异
\item \textbf{RQ4:} How does \shortname{} perform in real-world, large-scale online deployment?  % 上线提升【todo】
\end{itemize}

\subsection{Experimental Setup}
% \subsubsection{Datasets}
% 为了训练Reasoner，我们抽取了真实的日志数据 from kuaishou e-commerce platform，22M 在过去三个月内成果转化的dormant users交互历史，约28Mitems，交互sequence截断到128，对Reasoner进行sft训练。for behavioral backbone，而后对14M 纯dormant user进行反事实推理，其结果输入用来训练。详细见Appendix \ref{app:data}。
% 为了测试\shortname{}的效果，我们构造了采样了XM users的交互数据，with a candidate product pool of X million items，来构造测试数据。

% 
\noindent\textbf{Datasets.}
\textls[-22]{We conducted experiments using large-scale real-world datasets collected from the Kuaishou e-commerce platform.
To train the Reasoner, we constructed a corpus comprising 22 million users who successfully transitioned from dormant to active status within the past three months, involving approximately 28 million distinct items. Interaction sequences were truncated to a maximum length of 128.
Subsequently, we utilized 14 million dormant users to generate counterfactual reasoning signals, which were employed to fine-tune the behavioral backbone.
For evaluation, we randomly sampled a test set of dormant users, yielding more than 7 million positive user–item interaction pairs.
More details are provided in Appendix~\ref{app:data}.}

% \subsubsection{Metrics}
% In offline experiments, the performance is evaluated based on Hit Rate~\cite{}. 此外，对于生成式的推荐基座，我们额外增加针对生成的SID 的train loss、Hit SID$_l$ 、MRR SID$_l$进行统计。
% \item Hit @ K: It measures the ratio of ground-truth items that appear within the top K retrieved items from the beam search of the behavior backbone, 在所有样本数中。
% \item Loss：展示公式~\eqref{eq:loss_gen}的结果。
% \item HIT SID$_l$ @ K：给定ground truth SID$_{<l}$时，召回top K个next level SID$_{l}$ token，有多少出现了真实的SID$_{l}$。
% \item MRR SID$_l$ @ K: 类似于HIT SID$_l$ @ K，给定ground truth SID$_{<l}$时，召回top K个next level SID$_{l}$ token，计算真实的SID$_{l}$的Mean Reciprocal Ranking。

%In offline experiments, we primarily evaluate recommendation quality using Hit Rate at cut-off $K$ (H@K)~\cite{pang2025gram, tang2025lrem, chen2025onesearch}. For generative backbones, we additionally monitor token-level metrics on the semantic ID (SID) sequences.
\noindent\textbf{Metrics.}
In offline experiments, we evaluated the following metrics  to validate the effectiveness of \shortname{}:
\begin{itemize}[leftmargin=*, nosep]
    \item \textbf{Hit Item@K (HI@K)}: Proportion of test instances in which the ground-truth item appears among the top-$ K $ candidates generated by the behavioral backbone via beam search.
    \item \textbf{Loss}: Training loss $ L_{\text{Gen}} $ defined in Eq.~\eqref{eq:loss_gen}.
    %\item \textbf{Hit SID$_l$@K (HS$_l$@K)}: given the ground-truth prefix of SID levels $\mathrm{SID}_{<l}$, we retrieve the top-$K$ candidates for the next-level token $\mathrm{SID}_l$ and measure the fraction of cases where the true $\mathrm{SID}_l$ appears in this set.
    \item \textbf{Hit SID$_l$@K (HS$_l$@K)}: Fraction of cases where the ground-truth $ \mathrm{SID}_l $ is ranked within the top-$ K $ generated candidates.
    \item \textbf{MRR SID$_l$@K (MS$_l$@K)}: Mean Reciprocal Rank of the ground-truth $ \mathrm{SID}_l $ over all test instances, truncated at rank $ K $ ..
    %uses the same retrieval procedure as Hit SID$_l$@K, but averages the reciprocal rank of the ground-truth $\mathrm{SID}_l$ within the top-$K$ list.

\end{itemize}

% \subsubsection{Models}
% 我们在实验中对比以下版本：
% * SASRec~\cite{kang2018sasrec}: employs self- attention mechanisms to capture long-term dependencies in user interaction sequence.
% * TIGER~\cite{rajput2023tiger}: 经典的生成式推荐工作，introduces semantic ID via RQ-VAE, 将序列拼接成SID sequences后输入encoder、decoder依照自回归生成推荐。
% * KuaiBase：快手电商平台in-house 全量使用的召回模型
% * BehBackbone：本文所提的Generative Behavioral Backbone在平台所有用户上训练后的模型。
% * \shortname{}: 本文所提的完整模型，即在BehBackbone基础上加入Reasoning Guidance Features进行微调，并对两个模块进行synergistic training后的behavioral backbone。
% Model implementation details can be found in Appendix \ref{app:exp_model}.
\noindent\textbf{Models.}
\textls[-10]{We compare the following models in our experiments: (1) \textbf{\textit{SASRec}}~\cite{kang2018sasrec} is a state-of-the-art sequential recommendation model that captures long-term dependencies within interaction sequences. (2) \textbf{\textit{TIGER}}~\cite{rajput2023tiger} is a pioneering generative recommendation framework, which introduces semantic IDs via RQ-VAE and generates target SIDs autoregressively. (3) \textbf{\textit{U2I}} \& \textbf{\textit{I2I}} are the major online retrieval methods deployed on the Kuaishou e-commerce platform, serving full-scale online traffic, and accounting for more than 80\% of online orders.  (4) \textbf{\textit{w/o Rea}} is the base version of our proposed generative backbone, without Reasoner intervention. (5) \textbf{\textit{\shortname{}}} is the complete framework proposed in this paper. }
%It fine-tunes the Behavior Backbone by incorporating reasoning guidance features and employing the synergistic training strategy (Section~\ref{sec:synergy}).
\begin{comment}
\begin{itemize}[leftmargin=*]
    \item \textbf{SASRec}~\cite{kang2018sasrec}: A state-of-the-art sequential recommendation model that employs self-attention mechanisms to capture long-term dependencies within user interaction sequences.
    \item \textbf{TIGER}~\cite{rajput2023tiger}: A pioneering generative recommendation framework. It introduces semantic IDs via RQ-VAE and formulates recommendation as a sequence-to-sequence generation task, predicting target SIDs autoregressively.
    \item \textbf{U2I} \& \textbf{I2I}: The production-tier retrieval methods currently deployed on the Kuaishou e-commerce platform, including both U2I-type and I2I-type methods. U2Is (two-tower based model) and I2Is (like Swing~\cite{yang2020large} and PDN~\cite{li2021path}) are the major online methods, serving full-scale online traffic, and providing over 80\% online exposures, clicks, and orders. % todo：介绍快手的召回概况；说明展示的两个基线的代表性（占比）serving full-scale online traffic
    
    \item \textbf{w/o Reasoner}: The base version of our proposed generative backbone, trained on the entire platform user population without the intervention of the Reasoner.
    \item \textbf{\shortname{}}: The complete framework proposed in this paper. It fine-tunes the BehBackbone by incorporating reasoning guidance features and employing the synergistic training strategy (Section~\ref{sec:synergy}).
\end{itemize}
\end{comment}
\textls[-10]{Detailed implementation settings for all models are provided in Appendix~\ref{app:exp_model}.}

\begin{table}[t]
  \centering
  \vspace{-3pt}
  \caption{Comparison of offline evaluation results on item hitrate (HI@K) on Kuaishou e-commerce platform.}
  \renewcommand{\arraystretch}{0.9} % 稍微增加一点行高，防止多行文字过于拥挤
  \vspace{-8pt}
  \setlength{\tabcolsep}{2.1pt} % 稍微减小列间距以适应新增的一列
  {
    \footnotesize
    % 定义列格式：新增第一列 l，其余保持不变
    \begin{tabular}{llcccccc}
    \toprule
    \textbf{Category} & \textbf{Models} & \textbf{HI@1}   & \textbf{HI@10}  & \textbf{HI@100} & \textbf{HI@500} & \textbf{HI@2000} & \textbf{Loss}  \\
    \midrule
    
    % 第一组：Offline Sequential
    \multirow{2}{*}{\shortstack[l]{\textbf{Offline} \\ \textbf{SeqRec}}} 
  &  SASRec & 1.86\% & 2.93\% & 5.11\% & 14.37\% & 23.18\% &  -     \\
   & TIGER & 4.73\% & 7.88\% & 14.86\% & 19.53\% & 29.49\% &  -        \\
    \midrule
    
    % 第二组：Online Retrieval
    \multirow{2}{*}{\shortstack[l]{\textbf{Production} \\ \textbf{Retrieval}}} 
    & I2I & 4.38\% & 8.12\% & 17.82\% & 21.68\% & 31.76\% & - \\
    & U2I &  3.19\% & 4.26\% & 6.95\% & 15.74\% & 27.66\% & - \\
    \midrule
    
    % 第三组：Our Generative
    \multirow{2}{*}{\shortstack[l]{\textbf{Ours}}} 
      & w/o Rea & 5.11\% & 10.26\% & 20.54\% & 31.26\% & 35.16\% & 3.2950 \\
      & \shortname{} & \textbf{10.93}\% & \textbf{14.89}\% & \textbf{24.07}\% & \textbf{33.98}\% & \textbf{37.13\%} & \textbf{3.0443}  \\
      
    \bottomrule
    \end{tabular}}%
  \label{tab:offline_main_hi}%
  \vspace{-10pt}
\end{table}%

% \begin{table}[t]
%   \centering
%   \vspace{-12pt}
%   \caption{Comparison of offline evaluation results on item hitrate (HI@K) on Kuaishou e-commerce platform.}
%   \renewcommand{\arraystretch}{0.8}
%   \vspace{-13pt}
%   \setlength{\tabcolsep}{5pt}
%   {
%     \footnotesize
%     \begin{tabular}{lcccccc}
%     \toprule
%     \textbf{Models} & \textbf{HI@1}   & \textbf{HI@10}  & \textbf{HI@100} & \textbf{HI@500} & \textbf{HI@2000} & \textbf{Loss}  \\
%     \midrule
%     \multicolumn{7}{c}{\textbf{Offline Sequential Recommendation Methods}} \\
%     \midrule
%     SASRec & 1.86\% & 2.93\% & 5.11\% & 14.37\% & 23.18\% &  -     \\
%     TIGER & 4.73\% & 7.88\% & 14.86\% & 19.53\% & 29.49\% &  -        \\
%     \midrule
%     \multicolumn{7}{c}{\textbf{Main Online Retrieval Methonds}} \\
%     \midrule
%     I2I & 4.38\% & 8.12\% & 17.82\% & 21.68\% & 31.76\% & - \\
%     U2I &  3.19\% & 4.26\% & 6.95\% & 15.74\% & 27.66\% & - \\
%     \midrule
%     \multicolumn{7}{c}{\textbf{Our Generative Methond}} \\
%     \midrule
%     w/o Reasoner & 5.11\% & 10.26\% & 20.54\% & 31.26\% & 35.16\% & 3.2950 \\
%     \shortname{} & \textbf{10.93}\% & \textbf{14.89}\% & \textbf{24.07}\% & \textbf{33.98}\% & \textbf{37.13\%} & \textbf{3.0443}  \\
%     \bottomrule
%     \end{tabular}}%
%   \label{tab:offline_main_hi}%
%   \vspace{-13pt}
% \end{table}%

\subsection{Offline Evaluation (RQ1)}
\label{sec:offline}
\textls[-12]{Table \ref{tab:offline_main_hi} and \ref{tab:offline_main_hs} show the overall offline performance on the dormant users across competitive baselines, covering both item-level and semantic ID-level metrics. The results indicate that \shortname{} significantly outperforms all baselines on both tasks. 
This demonstrates its capability to effectively integrate collaborative signals with structured semantic knowledge via functional-role trajectory reasoning:}
% We detail our analysis against different baselines as follows:}
% The results indicate that \shortname{} significantly outperforms all the baselines by a large margin on both item ranking and semantic ID generation, demonstrating its ability to effectively integrate collaborative signals and structured semantic knowledge through functional-role trajectory reasoning. We analyze the effectiveness of our \shortname{} against different baselines as follows:}

First, our \shortname{} consistently achieves significant improvements across various evaluation metrics, demonstrating the effectiveness of our two model co-training strategy, i.e., world knowledge reasoning injection and collaborative behavior reflection. As the dormant users lack sufficient feedback signals, modeling only user-item interactions (as in baseline methods) exacerbates the data loop problem and leads to poor performance.  In contrast, \shortname{} generates high-quality candidates through world knowledge-based reasoning, thereby enabling more accurate recommendations.

Second, our \shortname{} beats non-generative methods on item hitrate, which proves the effectiveness of generative paradigm. Different from non-generative methods which directly select items from the entire massive candidate set, 
%generative methods split the whole candidate set into L layers by semantic ID, and generate items layer by layer. 
generative methods decompose the candidate space into $ L $ hierarchical layers based on semantic IDs and generate items layer by layer.
At each layer, the number of candidates decreases exponentially, resulting in a smoother and easier learning process. Consequently, generative methods achieve significant improvements, particularly at small cutoff values of $ K $ .

%In each layer, the number of candidates decrease exponentially, leading to smoother and easier learning process. Thus, generative methods presents significant improvements, especially on the small cutoff K values.

%Finally, adding the functional-role-guided reasoning also performs better than other generative methods (TIGER and w/o Reasoner) across 
\textls[-10]{Finally, incorporating functional-role-guided reasoning further improves performance over other generative methods (TIGER and w/o Reasoner) on both item Hitrate and semantic ID Hitrate \& MRR. 
%It indicates that behavior interactions are not enough for dormant users. 
This indicates that behavioral interactions alone are insufficient for dormant users.
%By interacting the Functional Role Trajectory, \shortname{} explicitly understands inherent static properties of each behavioral item in a common knowledge manner. 
By modeling the Functional Role Trajectory, \shortname{} explicitly captures the inherent static properties of each interacted item in a common sense knowledge framework.
As a result, \shortname{} effectively captures the preferences of dormant users.
%Thus, \shortname{} works well on capturing dormant users' preference. 
Moreover, compared with w/o Reasoner and RoleGen, the performance gain is primarily observed at SID $ _1 $ and SID $ _2 $ , as our co-training strategy mainly leverages statistics from the first two layers of SIDs generated by the Reasoner (see Section~\ref{sec:synergy} for details).}

\begin{table}[t]
  \centering
    % \vspace{-0pt}
  \caption{Comparison of offline evaluation results on SID Hitrate \& MRR across Generative Methods.}
  \renewcommand{\arraystretch}{0.85}
   \vspace{-8pt}
  \setlength{\tabcolsep}{2.6pt}
  {
    \footnotesize
    \begin{tabular}{lcccccc}
    \toprule
    \textbf{Models}  & \textbf{HS$\_1$@1} & \textbf{HS$\_2$@1} & \textbf{HS$\_3$@1} & \textbf{MS$\_1$@10} & \textbf{MS$\_2$@10} & \textbf{MS$\_3$@10} \\
    \midrule
    TIGER & 19.83\% & 33.47\% & 50.24\% & 27.35\% & 46.59\% & 60.16\%         \\
    w/o Reasoner & 24.51\% & 38.09\% & 56.32\% & 31.55\% & 51.43\% & 64.93\% \\
    \shortname{} & \textbf{30.12\%} & \textbf{41.77\%} & \textbf{56.41\%} & \textbf{38.72\%} & \textbf{53.35\%} & \textbf{65.28\%}  \\
    \bottomrule
    \end{tabular}}%
  \label{tab:offline_main_hs}%
    \vspace{-13pt}
\end{table}%

\subsection{Ablation Study (RQ2)}
To evaluate the effectiveness of each component of \shortname{}, we conduct the ablation study for Generative Behavioral Backbone and Conversion Trajectory Reasoner.

First, for Generative Behavioral Backbone part, Section \ref{sec:offline} presents significant improvement by incorporating trajectory reasoning enhancement. For dormant users with limited behavioral signals, counterfactual functional role inference provides generalized and novel supplements, and alievates the self-reinforcing loop.

\begin{figure}[!t]
    \centering
    \vspace{-10pt}
    \includegraphics[width=0.45\textwidth, trim=-0 0 -0 0, clip]{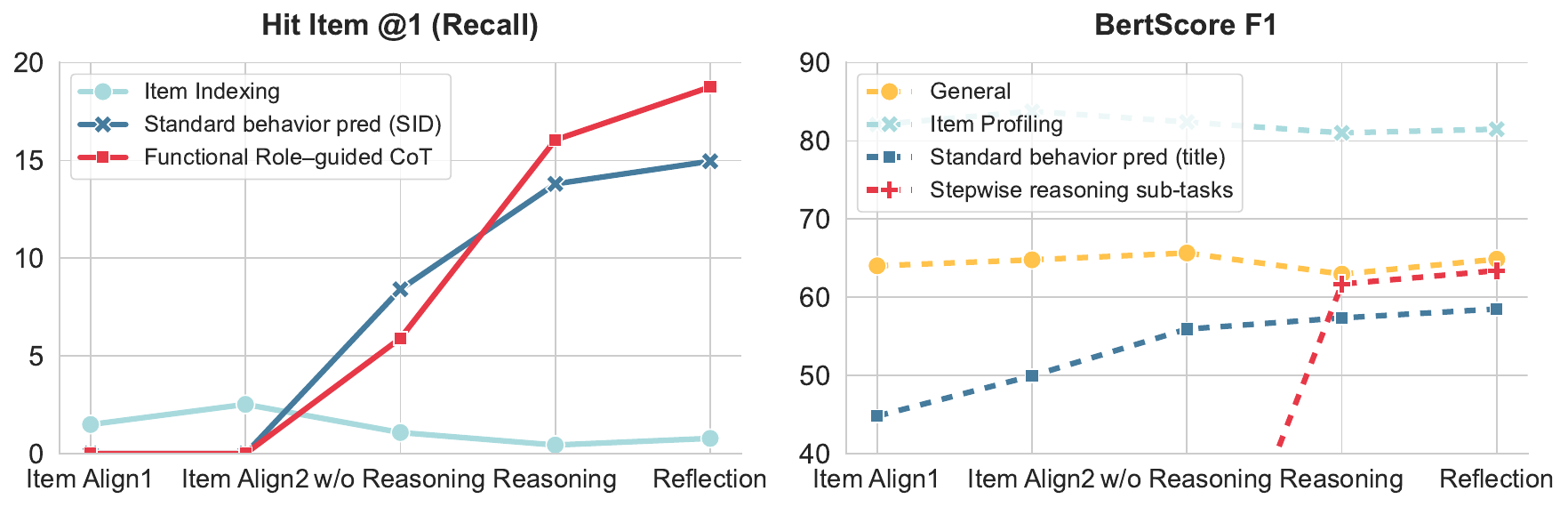}
    \vspace{-8pt}
    \caption{Ablation study of \shortname{}.} 
    % The results demonstrate that for both item prediction and token generation tasks, every component (including SID alignment, Functional Role Reasoning, and the Reflection Mechanism) contributes significantly to the final improvements.
    \vspace{-12pt}
    \label{fig:ablation}
\end{figure}

Second, for the training of the Conversion Trajectory Reasoner, we dissect the impact of four key components as shown in Fig. \ref{fig:ablation}: 
SID Alignment (comprising \textbf{Item Align 1} on the global item set and \textbf{Item Align 2} on the selected high-quality subset, see Sec.~\ref{app:data}), 
standard next item prediction (\textbf{w/o Reasoning}), 
functional-role reasoning enhance (\textbf{Reasoning}), 
and the reflection mechanism guided by collaborative feedback (\textbf{Reflection}). As the evaluation tasks include both item predictions and token generation (e.g. item title and item profile generation), we adopt hit item and BertScore~\cite{liu2025onerecthink} for them, respectively. From Fig. \ref{fig:ablation}, we can see that each componet of the Conversion Trajectory Reasoner contributes to the dormant user modeling, as the performance increases with any of the views added. We can see that for item prediction task, Functional Role guided CoT outperforms Standard SID prediction when adding supervised trajectory reasoning, which indicates that the trajectory reasoner could help to capture dormant user intents. Moreover, the general reasoning ability keeps well.

% 展示Reasoner不同SFT阶段的ckpt在各个任务上的差距，证明每个阶段的能力和必要性；Behavior backbone的contribution见上一节
% 对于生成文字类的SFT任务，使用BertScore~\cite{liu2025onerecthink}评估生成质量；对于生成SID的任务，使用Hit Item指标
% Item Align的两个阶段分别是指全量Item训练和精选Item训练；w/o Reasoning是指使用标准的seq -> next item的sft任务训练，即没有中间Functional role思考的过程；reflection就是指加进onerec再进行一轮训练

\begin{figure}[!t]
    \centering
    % \vspace{-16pt}
    \includegraphics[width=0.40\textwidth, trim=-0 0 -0 30, clip]{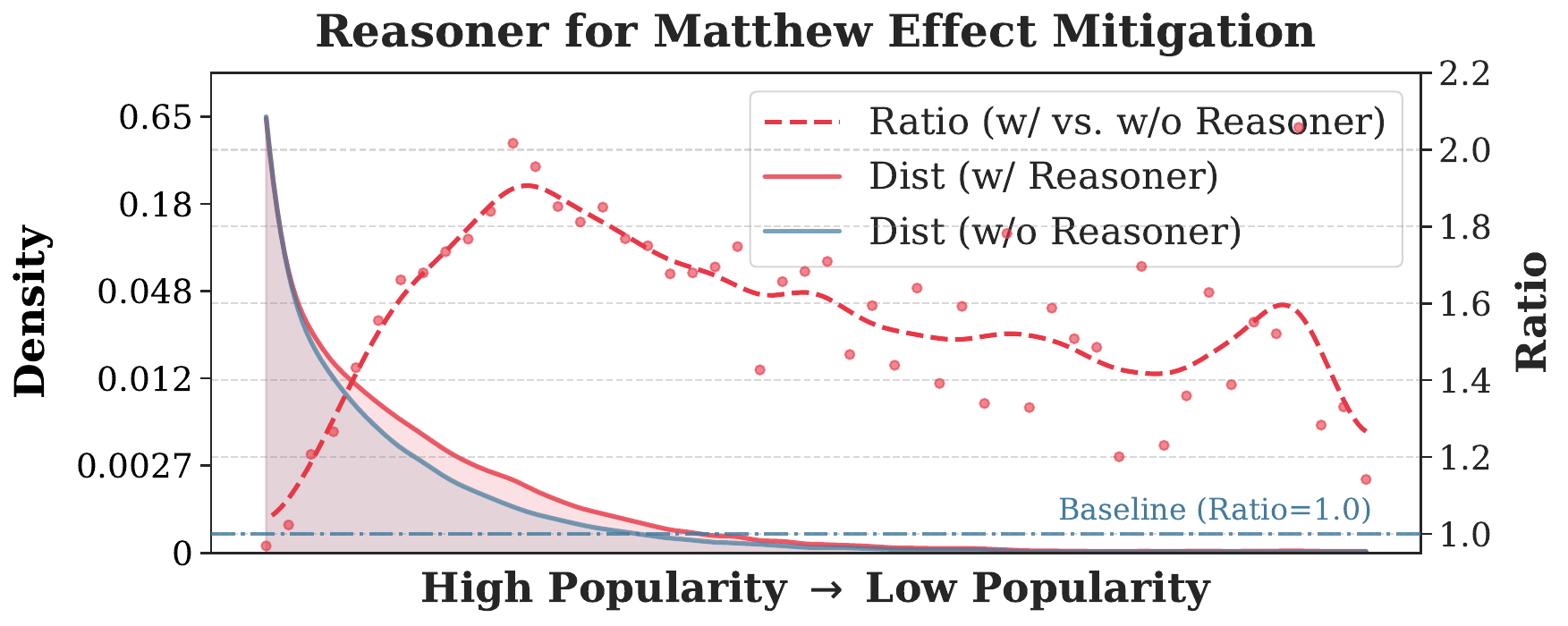}
    \vspace{-10pt}
    \caption{Analysis of Matthew effect mitigation. The exposure ratio rises significantly above 1.0 for long-tail items.
    % in the low-popularity region
    % , indicating that the Reasoner successfully enhances the exploration of long-tail items.
    }
    \vspace{-4pt}
    \label{fig:matthew}
\end{figure}

\begin{figure}[!t]
    \centering
    % \vspace{-6pt}
    \includegraphics[width=0.48\textwidth, trim=-0 6 -0 0, clip]{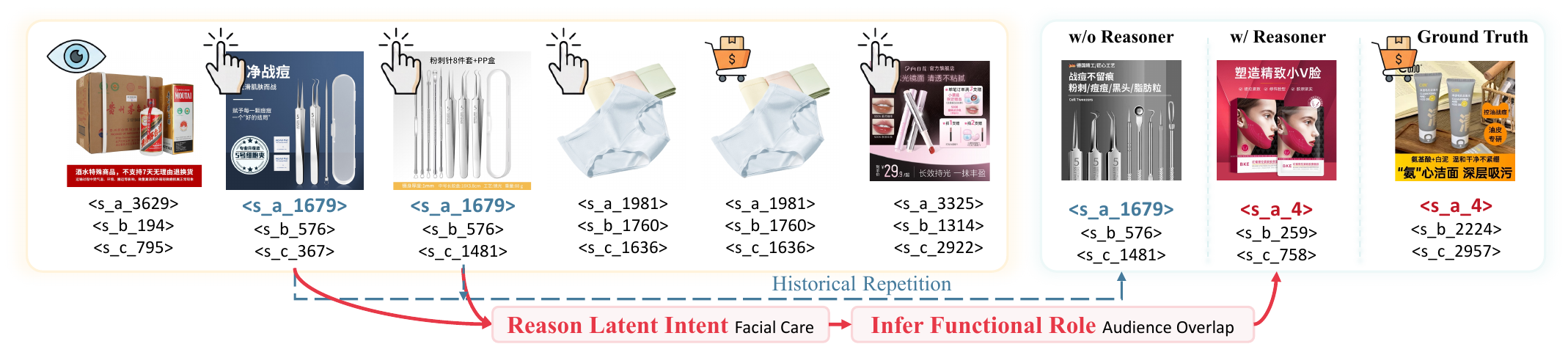}
    \vspace{-16pt}
    \caption{Case study on intent generalization. Unlike the baseline which suffers from interest collapse, \shortname{} leverages “Audience Overlap” functional roles to break the self-reinforcing loop, achieving precise Level-1 SID prediction.}
    \vspace{-5pt}
    \label{fig:case}
\end{figure}

\subsection{Generalization Analysis \& Case Study (RQ3)}
\label{sec:case}
% popularity分布证明了缓解了the Matthew effect，OOD test结果证明了mitigating both filter bubbles for dormant users，两者结合起来共同证明了我们能够跳出恶性循环。
% case Study（onerec差，我们的好，并给出Functional Role示例）

In this section, we develop the following experiments to show that world knowledge reasoning yields novel recommendations and exhibits stronger generalization on long-tail interactions.
%provides novel results, and shows stronger generalization on long-tail interactions.

\textls[-15]{First, we partition items into several buckets based on popularity, and compute the item numbers ratio (w/ Reasoner vs. w/o Reasoner) in each bucket. The results are shown in Fig. \ref{fig:matthew}. 
The results indicate that incorporating the Reasoner significantly boosts the coverage of long-tail items while reducing the concentration on the most popular bucket.
This confirms that the Reasoner effectively alleviates the Matthew Effect.
% This indicates that incorporating the world knowledge Reasoner significantly improves the representation of long-tail item candidates, except for the most popular bucket. Thus, the Reasoner helps mitigate the Matthew Effect.
}

\textls[-12]{Then, we adopt out-of-distribution (OOD) test on dormant users. 
We define an interaction as OOD if the target item's top-level semantic category (SID$_{1}$) has never appeared in the user’s historical sequence.
% The results are shown in Fig. \ref{tab:generalized_ood}. 
%We observe that Reasoner provides strong generalization items especially on OOD scene.
As the results in Table \ref{tab:generalized_ood}, we observe that the Reasoner generates significantly more generalizable recommendations, especially in the OOD scenario.}

\textls[-12]{To provide a more intuitive comparison, we present a representative case study in Figure \ref{fig:case}. Historically, the user's interactions were primarily focused on intimate apparel and cosmetics. However, the ground truth reveals a shift in the user's latent intent towards facial care products. The “w/o Reasoner” baseline, constrained by historical patterns, myopically recommends repeated items and fails to escape the self-reinforcing feedback loop. In contrast, \shortname{} successfully infers the latent intent and recommends a face mask. This demonstrates that \shortname{} achieves superior generalization aligned with the user's evolving intent, effectively breaking the filter bubble.}

\begin{table}[t]
  \centering
  \vspace{-10pt}
  \caption{OOD test on functional role reasoning. The world knowledge shows more generalization on OOD data, which helps provide novel items.}
  \vspace{-8pt}
  \setlength{\tabcolsep}{2.6pt}
  {
    % \footnotesize
    \begin{tabular}{lcccc}
    \toprule
    \textbf{Models}  & \textbf{HS$\_1$@10} & \textbf{HS$\_1$@20} & \textbf{HS$\_2$@10} & \textbf{HS$\_2$@20} \\
    \midrule
    w/o Reasoner & 2.06\% & 4.33\% & 10.97\% & 15.88\% \\
    \shortname{} & \textbf{3.12\%} & \textbf{5.25\%} & \textbf{20.53\%} & \textbf{28.48\%} \\
    \bottomrule
    \end{tabular}}%
  \label{tab:generalized_ood}%
  \vspace{-5pt}
\end{table}%

% \begin{figure}[t]
%   \centering
%   % 左图
%   \begin{minipage}[b]{0.3\linewidth}
%     \includegraphics[width=\linewidth]{pic/matthew_effect_analysis.pdf} % 
%   \end{minipage}
%   \hfill % 这里的 \hfill 是关键，用于撑开两张图
%   % 右图
%   \begin{minipage}[b]{0.68\linewidth}
%     \includegraphics[width=\linewidth]{pic/case2.pdf} % 替换为你的文件名
%   \end{minipage}
  
%   % \vspace{-2mm} % 可选：调整图与 Caption 的间距
%   \caption{(Left) Analysis of Matthew effect mitigation. The exposure ratio rises significantly above 1.0 in the low-popularity region, indicating that the Reasoner successfully enhances the exploration of long-tail items. (Right) Case study on intent generalization. The baseline merely retrieves historical items, leading to interest collapse. In contrast, \shortname{} identifies the user intent and infers the "Audience Overlap" functional role to generalize recommendations. The generated candidate achieves accurate Level-1 SID alignment ($\texttt{<s\_a\_4>}$) with the ground truth, effectively breaking the self-reinforcing feedback loop.}
%   \label{fig:two_images}
% \end{figure}

% 图的大小会再调

\begin{table}[t]
  \centering
  %\vspace{-10pt}
  \caption{Online A/B testing results improvements over the production baseline on dormant users (p < 0.05).}
  \renewcommand{\arraystretch}{0.9}
  \vspace{-5pt}
  \setlength{\tabcolsep}{2.6pt}
  {
    % \footnotesize
    \begin{tabular}{c|c|c|c}
    \toprule
    \textbf{Model}  & \textbf{DAC} & \textbf{CTCVR} & \textbf{Order Volume} \\
    \midrule
    \shortname{} & \textbf{+5.57\%} & \textbf{+5.38\%} & \textbf{+6.71\%} \\
    \bottomrule
    \end{tabular}}%
  \label{tab:online_ab}%
  \vspace{-5pt}
\end{table}%

% \begin{table}[t]
%   \centering
%   \vspace{-0pt}
%   \caption{Online A/B test results (p < 0.05) on dormant users.}
%   %\renewcommand{\arraystretch}{0.9}
%   \vspace{-0pt}
%   \setlength{\tabcolsep}{2.6pt}
%   {
%     \footnotesize
%     \begin{tabular}{c|c|c|c}
%     \toprule
%     \textbf{Model}  & \textbf{DAC} & \textbf{CTCVR} & \textbf{Goods Orders} \\
%     \midrule
%     Baseline & 0 & 0 & 0 \\
%     \shortname{} & \textbf{+5.57\%} & \textbf{+5.38\%} & \textbf{+6.71\%} \\
%     \bottomrule
%     \end{tabular}}%
%   \label{tab:online_ab}%
%   \vspace{-11pt}
% \end{table}%

\subsection{Online Evaluation (RQ4)}
We further validate the practical efficacy of \shortname{} by online A/B test. We allocate 10\% of the traffic over several weeks for A/B testing, and the results are shown in Table \ref{tab:online_ab}. Specifically, \shortname{} achieves a significant \textbf{+5.57\%} Daily Active Customers (DAC) and \textbf{+6.71\%} Goods Orders improvement in real-world industrial scenarios. These results confirm that \shortname{} provides high-quality, generalized, and novel candidates for dormant users, and breaks the self-reinforce loop, which leads to a health recommendation ecosystem on Kuaishou e-commerce platform.

% Online deployment 见Section~\ref{sec:deployment}；简单总结涨，以及涨的缘由

\section{Conclusion}
\textls[-12]{This paper proposes \shortname{}, a novel reasoning enhanced retriever to re-engage dormant users. 
In contrast to previous methods that focus on isolated point-wise item prediction, \shortname{} first extracts functional role trajectories via structured CoT, subsequently integrating the inferred user intent into a collaborative filtering (CF) backbone.
Finally, we establish a closed-loop mechanism “Reasoning-Execution-Feedback-Reflection” to synergize the semantic Reasoner with the CF-based Behavioral Backbone.
Extensive offline and online experiments verify the effectiveness of \shortname{}.
}

% \begin{acks}
% \end{acks}

\bibliographystyle{ACM-Reference-Format}
\bibliography{ref}

@inproceedings{hua2023tutorial,
  title={Tutorial on large language models for recommendation},
  author={Hua, Wenyue and Li, Lei and Xu, Shuyuan and Chen, Li and Zhang, Yongfeng},
  booktitle={Proceedings of the 17th ACM Conference on Recommender Systems},
  pages={1281--1283},
  year={2023}
}

@article{huang2024large,
  title={Large Language Model Interaction Simulator for Cold-Start Item Recommendation},
  author={Huang, Feiran and Yang, Zhenghang and Jiang, Junyi and Bei, Yuanchen and Zhang, Yijie and Chen, Hao},
  journal={arXiv preprint arXiv:2402.09176},
  year={2024}
}

@inproceedings{yang2024common,
  title={Common sense enhanced knowledge-based recommendation with large language model},
  author={Yang, Shenghao and Ma, Weizhi and Sun, Peijie and Zhang, Min and Ai, Qingyao and Liu, Yiqun and Cai, Mingchen},
  booktitle={International Conference on Database Systems for Advanced Applications},
  pages={381--390},
  year={2024},
  organization={Springer}
}

@inproceedings{cui2025comprehending,
  title={Comprehending knowledge graphs with large language models for recommender systems},
  author={Cui, Ziqiang and Weng, Yunpeng and Tang, Xing and Lyu, Fuyuan and Liu, Dugang and He, Xiuqiang and Ma, Chen},
  booktitle={Proceedings of the 48th international ACM SIGIR conference on research and development in information retrieval},
  pages={1229--1239},
  year={2025}
}

@article{lin2025can,
  title={How can recommender systems benefit from large language models: A survey},
  author={Lin, Jianghao and Dai, Xinyi and Xi, Yunjia and Liu, Weiwen and Chen, Bo and Zhang, Hao and Liu, Yong and Wu, Chuhan and Li, Xiangyang and Zhu, Chenxu and others},
  journal={ACM Transactions on Information Systems},
  volume={43},
  number={2},
  pages={1--47},
  year={2025},
  publisher={ACM New York, NY}
}

@article{chen2024large,
  title={When large language models meet personalization: Perspectives of challenges and opportunities},
  author={Chen, Jin and Liu, Zheng and Huang, Xu and Wu, Chenwang and Liu, Qi and Jiang, Gangwei and Pu, Yuanhao and Lei, Yuxuan and Chen, Xiaolong and Wang, Xingmei and others},
  journal={World Wide Web},
  volume={27},
  number={4},
  pages={42},
  year={2024},
  publisher={Springer}
}

@article{wu2024survey,
  title={A survey on large language models for recommendation},
  author={Wu, Likang and Zheng, Zhi and Qiu, Zhaopeng and Wang, Hao and Gu, Hongchao and Shen, Tingjia and Qin, Chuan and Zhu, Chen and Zhu, Hengshu and Liu, Qi and others},
  journal={World Wide Web},
  volume={27},
  number={5},
  pages={60},
  year={2024},
  publisher={Springer}
}

@misc{liu2025llmal,
      title={LLM-Alignment Live-Streaming Recommendation}, 
      author={Yueyang Liu and Jiangxia Cao and Shen Wang and Shuang Wen and Xiang Chen and Xiangyu Wu and Shuang Yang and Zhaojie Liu and Kun Gai and Guorui Zhou},
      year={2025},
      eprint={2504.05217},
      archivePrefix={arXiv},
      primaryClass={cs.IR},
      url={https://arxiv.org/abs/2504.05217}, 
}

@inproceedings{Gu2025R4ec,
author = {Gu, Hao and Zhong, Rui and Xia, Yu and Yang, Wei and Lu, Chi and Jiang, Peng and Gai, Kun},
title = {R4ec: A Reasoning, Reflection, and Refinement Framework for Recommendation Systems},
year = {2025},
isbn = {9798400713644},
publisher = {Association for Computing Machinery},
address = {New York, NY, USA},
url = {https://doi.org/10.1145/3705328.3748068},
doi = {10.1145/3705328.3748068},
booktitle = {Proceedings of the Nineteenth ACM Conference on Recommender Systems},
pages = {411–421},
numpages = {11},
location = {
},
series = {RecSys '25}
}

@article{ge2025llm,
  title={Llm-enhanced composed image retrieval: An intent uncertainty-aware linguistic-visual dual channel matching model},
  author={Ge, Hongfei and Jiang, Yuanchun and Sun, Jianshan and Yuan, Kun and Liu, Yezheng},
  journal={ACM Transactions on Information Systems},
  volume={43},
  number={2},
  pages={1--30},
  year={2025},
  publisher={ACM New York, NY}
}

@inproceedings{cheng2025poi,
  title={Poi-enhancer: An llm-based semantic enhancement framework for poi representation learning},
  author={Cheng, Jiawei and Wang, Jingyuan and Zhang, Yichuan and Ji, Jiahao and Zhu, Yuanshao and Zhang, Zhibo and Zhao, Xiangyu},
  booktitle={Proceedings of the AAAI conference on artificial intelligence},
  volume={39},
  number={11},
  pages={11509--11517},
  year={2025}
}

@article{sun2025llmser,
  title={LLMSeR: Enhancing Sequential Recommendation via LLM-based Data Augmentation},
  author={Sun, Yuqi and Liu, Qidong and Zhu, Haiping and Tian, Feng},
  journal={arXiv preprint arXiv:2503.12547},
  year={2025}
}

@article{liu2025filterllm,
  title={FilterLLM: Text-To-Distribution LLM for Billion-Scale Cold-Start Recommendation},
  author={Liu, Ruochen and Chen, Hao and Bei, Yuanchen and Zhou, Zheyu and Chen, Lijia and Shen, Qijie and Huang, Feiran and Karray, Fakhri and Wang, Senzhang},
  journal={arXiv preprint arXiv:2502.16924},
  year={2025}
}

@inproceedings{liu2025llm-esr,
author = {Liu, Qidong and Wu, Xian and Wang, Yejing and Zhang, Zijian and Tian, Feng and Zheng, Yefeng and Zhao, Xiangyu},
title = {LLM-ESR: large language models enhancement for long-tailed sequential recommendation},
year = {2025},
isbn = {9798331314385},
publisher = {Curran Associates Inc.},
address = {Red Hook, NY, USA},
booktitle = {Proceedings of the 38th International Conference on Neural Information Processing Systems},
articleno = {839},
numpages = {27},
location = {Vancouver, BC, Canada},
series = {NIPS '24}
}

@inproceedings{Wei2024LLMRec,
author = {Wei, Wei and Ren, Xubin and Tang, Jiabin and Wang, Qinyong and Su, Lixin and Cheng, Suqi and Wang, Junfeng and Yin, Dawei and Huang, Chao},
title = {LLMRec: Large Language Models with Graph Augmentation for Recommendation},
year = {2024},
isbn = {9798400703713},
publisher = {Association for Computing Machinery},
address = {New York, NY, USA},
url = {https://doi.org/10.1145/3616855.3635853},
doi = {10.1145/3616855.3635853},
booktitle = {Proceedings of the 17th ACM International Conference on Web Search and Data Mining},
pages = {806–815},
numpages = {10},
location = {Merida, Mexico},
series = {WSDM '24}
}

@article{rajput2023recommender,
  title={Recommender systems with generative retrieval},
  author={Rajput, Shashank and Mehta, Nikhil and Singh, Anima and Hulikal Keshavan, Raghunandan and Vu, Trung and Heldt, Lukasz and Hong, Lichan and Tay, Yi and Tran, Vinh and Samost, Jonah and others},
  journal={Advances in Neural Information Processing Systems},
  volume={36},
  pages={10299--10315},
  year={2023}
}

@article{deng2025onerec,
  title={Onerec: Unifying retrieve and rank with generative recommender and iterative preference alignment},
  author={Deng, Jiaxin and Wang, Shiyao and Cai, Kuo and Ren, Lejian and Hu, Qigen and Ding, Weifeng and Luo, Qiang and Zhou, Guorui},
  journal={arXiv preprint arXiv:2502.18965},
  year={2025}
}

@inproceedings{wang2025generative,
  title={Generative large recommendation models: Emerging trends in llms for recommendation},
  author={Wang, Hao and Guo, Wei and Zhang, Luankang and Chin, Jin Yao and Ye, Yufei and Guo, Huifeng and Liu, Yong and Lian, Defu and Tang, Ruiming and Chen, Enhong},
  booktitle={Companion Proceedings of the ACM on Web Conference 2025},
  pages={49--52},
  year={2025}
}

@article{zhang2025cold,
  title={Cold-start recommendation towards the era of large language models (llms): A comprehensive survey and roadmap},
  author={Zhang, Weizhi and Bei, Yuanchen and Yang, Liangwei and Zou, Henry Peng and Zhou, Peilin and Liu, Aiwei and Li, Yinghui and Chen, Hao and Wang, Jianling and Wang, Yu and others},
  journal={arXiv preprint arXiv:2501.01945},
  year={2025}
}

@inproceedings{han2025mtgr,
  title={Mtgr: Industrial-scale generative recommendation framework in meituan},
  author={Han, Ruidong and Yin, Bin and Chen, Shangyu and Jiang, He and Jiang, Fei and Li, Xiang and Ma, Chi and Huang, Mincong and Li, Xiaoguang and Jing, Chunzhen and others},
  booktitle={Proceedings of the 34th ACM International Conference on Information and Knowledge Management},
  pages={5731--5738},
  year={2025}
}

@article{huang2025impact,
  title={The Impact of Personalized Recommendation Systems on Consumer Purchase Decisions Under Data Law Frameworks: An Empirical Study Based on E-Commerce User Behavior Data},
  author={Huang, Silong and Liu, Zichen},
  journal={Journal of Organizational and End User Computing (JOEUC)},
  volume={37},
  number={1},
  pages={1--28},
  year={2025},
  publisher={IGI Global Scientific Publishing}
}

@misc{yoo2025continuallowrankadaptersllmbased,
      title={Continual Low-Rank Adapters for LLM-based Generative Recommender Systems}, 
      author={Hyunsik Yoo and Ting-Wei Li and SeongKu Kang and Zhining Liu and Charlie Xu and Qilin Qi and Hanghang Tong},
      year={2025},
      eprint={2510.25093},
      archivePrefix={arXiv},
      primaryClass={cs.LG},
      url={https://arxiv.org/abs/2510.25093}, 
}

@misc{lee2025capturinguserinterestsdata,
      title={Capturing User Interests from Data Streams for Continual Sequential Recommendation}, 
      author={Gyuseok Lee and Hyunsik Yoo and Junyoung Hwang and SeongKu Kang and Hwanjo Yu},
      year={2025},
      eprint={2506.07466},
      archivePrefix={arXiv},
      primaryClass={cs.IR},
      url={https://arxiv.org/abs/2506.07466}, 
}

@inproceedings{chang2023latent,
  title={Latent user intent modeling for sequential recommenders},
  author={Chang, Bo and Karatzoglou, Alexandros and Wang, Yuyan and Xu, Can and Chi, Ed H and Chen, Minmin},
  booktitle={Companion Proceedings of the ACM Web Conference 2023},
  pages={427--431},
  year={2023}
}

@article{kutlimuratov2022modeling,
  title={Modeling and applying implicit dormant features for recommendation via clustering and deep factorization},
  author={Kutlimuratov, Alpamis and Abdusalomov, Akmalbek Bobomirzaevich and Oteniyazov, Rashid and Mirzakhalilov, Sanjar and Whangbo, Taeg Keun},
  journal={Sensors},
  volume={22},
  number={21},
  pages={8224},
  year={2022},
  publisher={MDPI}
}

@article{qu2025tokenrec,
  title={Tokenrec: Learning to tokenize id for llm-based generative recommendations},
  author={Qu, Haohao and Fan, Wenqi and Zhao, Zihuai and Li, Qing},
  journal={IEEE Transactions on Knowledge and Data Engineering},
  year={2025},
  publisher={IEEE}
}

@inproceedings{luo2025qarm_sid,
  title={Qarm: Quantitative alignment multi-modal recommendation at kuaishou},
  author={Luo, Xinchen and Cao, Jiangxia and Sun, Tianyu and Yu, Jinkai and Huang, Rui and Yuan, Wei and Lin, Hezheng and Zheng, Yichen and Wang, Shiyao and Hu, Qigen and others},
  booktitle={Proceedings of the 34th ACM International Conference on Information and Knowledge Management},
  pages={5915--5922},
  year={2025}
}

@article{tang2025lrem,
  title={Large Reasoning Embedding Models: Towards Next-Generation Dense Retrieval Paradigm},
  author={Tang, Jianting and Li, Dongshuai and Wen, Tao and Lv, Fuyu and Ou, Dan and Xu, Linli},
  journal={arXiv preprint arXiv:2510.14321},
  year={2025}
}

@article{zeng2025grm,
  title={Optimizing Generative Ranking Relevance via Reinforcement Learning in Xiaohongshu Search},
  author={Zeng, Ziyang and Jing, Heming and Chen, Jindong and Li, Xiangli and Liu, Hongyu and He, Yixuan and Li, Zhengyu and Sun, Yige and Xie, Zheyong and Yang, Yuqing and others},
  journal={arXiv preprint arXiv:2512.00968},
  year={2025}
}

@article{ye2025align3gr,
  title={Align$^{3}$ GR: Unified Multi-Level Alignment for LLM-based Generative Recommendation},
  author={Ye, Wencai and Sun, Mingjie and Chen, Shuhang and Wu, Wenjin and Jiang, Peng},
  journal={arXiv preprint arXiv:2511.11255},
  year={2025}
}

@article{yi2025recgpt,
  title={Recgpt technical report},
  author={Yi, Chao and Chen, Dian and Guo, Gaoyang and Tang, Jiakai and Wu, Jian and Yu, Jing and Zhang, Mao and Dai, Sunhao and Chen, Wen and Yang, Wenjun and others},
  journal={arXiv preprint arXiv:2507.22879},
  year={2025}
}

@article{liu2025onerecthink,
  title={Onerec-think: In-text reasoning for generative recommendation},
  author={Liu, Zhanyu and Wang, Shiyao and Wang, Xingmei and Zhang, Rongzhou and Deng, Jiaxin and Bao, Honghui and Zhang, Jinghao and Li, Wuchao and Zheng, Pengfei and Wu, Xiangyu and others},
  journal={arXiv preprint arXiv:2510.11639},
  year={2025}
}

@article{peng2023gpt4data,
  title={Instruction tuning with gpt-4},
  author={Peng, Baolin and Li, Chunyuan and He, Pengcheng and Galley, Michel and Gao, Jianfeng},
  journal={arXiv preprint arXiv:2304.03277},
  year={2023}
}

@article{luo2025empirical,
  title={An empirical study of catastrophic forgetting in large language models during continual fine-tuning},
  author={Luo, Yun and Yang, Zhen and Meng, Fandong and Li, Yafu and Zhou, Jie and Zhang, Yue},
  journal={IEEE Transactions on Audio, Speech and Language Processing},
  year={2025},
  publisher={IEEE}
}

@article{zhou2025onerecv2,
  title={Onerec-v2 technical report},
  author={Zhou, Guorui and Hu, Hengrui and Cheng, Hongtao and Wang, Huanjie and Deng, Jiaxin and Zhang, Jinghao and Cai, Kuo and Ren, Lejian and Ren, Lu and Yu, Liao and others},
  journal={arXiv preprint arXiv:2508.20900},
  year={2025}
}

@inproceedings{kang2018sasrec,
  title={Self-attentive sequential recommendation},
  author={Kang, Wang-Cheng and McAuley, Julian},
  booktitle={2018 IEEE international conference on data mining (ICDM)},
  pages={197--206},
  year={2018},
  organization={IEEE}
}

@article{rajput2023tiger,
  title={Recommender systems with generative retrieval},
  author={Rajput, Shashank and Mehta, Nikhil and Singh, Anima and Hulikal Keshavan, Raghunandan and Vu, Trung and Heldt, Lukasz and Hong, Lichan and Tay, Yi and Tran, Vinh and Samost, Jonah and others},
  journal={Advances in Neural Information Processing Systems},
  volume={36},
  pages={10299--10315},
  year={2023}
}

@inproceedings{zheng2024llamafactory,
  title={LlamaFactory: Unified Efficient Fine-Tuning of 100+ Language Models},
  author={Yaowei Zheng and Richong Zhang and Junhao Zhang and Yanhan Ye and Zheyan Luo and Zhangchi Feng and Yongqiang Ma},
  booktitle={Proceedings of the 62nd Annual Meeting of the Association for Computational Linguistics (Volume 3: System Demonstrations)},
  address={Bangkok, Thailand},
  publisher={Association for Computational Linguistics},
  year={2024},
  url={http://arxiv.org/abs/2403.13372}
}

@article{yang2020large,
  title={Large scale product graph construction for recommendation in e-commerce},
  author={Yang, Xiaoyong and Zhu, Yadong and Zhang, Yi and Wang, Xiaobo and Yuan, Quan},
  journal={arXiv preprint arXiv:2010.05525},
  year={2020}
}

@inproceedings{li2021path,
  title={Path-based deep network for candidate item matching in recommenders},
  author={Li, Houyi and Chen, Zhihong and Li, Chenliang and Xiao, Rong and Deng, Hongbo and Zhang, Peng and Liu, Yongchao and Tang, Haihong},
  booktitle={Proceedings of the 44th International ACM SIGIR Conference on Research and Development in Information Retrieval},
  pages={1493--1502},
  year={2021}
}

@inproceedings{zhang2025llmtreerec,
  title={Llmtreerec: Unleashing the power of large language models for cold-start recommendations},
  author={Zhang, Wenlin and Wu, Chuhan and Li, Xiangyang and Wang, Yuhao and Dong, Kuicai and Wang, Yichao and Dai, Xinyi and Zhao, Xiangyu and Guo, Huifeng and Tang, Ruiming},
  booktitle={Proceedings of the 31st International Conference on Computational Linguistics},
  pages={886--896},
  year={2025}
}

@inproceedings{tao2022sminet,
  title={SMINet: State-aware multi-aspect interests representation network for cold-start users recommendation},
  author={Tao, Wanjie and Li, Yu and Li, Liangyue and Chen, Zulong and Wen, Hong and Chen, Peilin and Liang, Tingting and Lu, Quan},
  booktitle={Proceedings of the AAAI conference on artificial intelligence},
  volume={36},
  number={8},
  pages={8476--8484},
  year={2022}
}

@inproceedings{huang2023aligning,
  title={Aligning distillation for cold-start item recommendation},
  author={Huang, Feiran and Wang, Zefan and Huang, Xiao and Qian, Yufeng and Li, Zhetao and Chen, Hao},
  booktitle={Proceedings of the 46th international ACM SIGIR conference on research and development in information retrieval},
  pages={1147--1157},
  year={2023}
}

@article{liu2023contrastive,
  title={Contrastive proxy kernel stein path alignment for cross-domain cold-start recommendation},
  author={Liu, Weiming and Zheng, Xiaolin and Su, Jiajie and Zheng, Longfei and Chen, Chaochao and Hu, Mengling},
  journal={IEEE Transactions on Knowledge and Data Engineering},
  volume={35},
  number={11},
  pages={11216--11230},
  year={2023},
  publisher={IEEE}
}

@inproceedings{zhu2021learning,
  title={Learning to warm up cold item embeddings for cold-start recommendation with meta scaling and shifting networks},
  author={Zhu, Yongchun and Xie, Ruobing and Zhuang, Fuzhen and Ge, Kaikai and Sun, Ying and Zhang, Xu and Lin, Leyu and Cao, Juan},
  booktitle={Proceedings of the 44th International ACM SIGIR Conference on Research and Development in Information Retrieval},
  pages={1167--1176},
  year={2021}
}

@article{xu2025c2lrec,
  title={C2lrec: causal contrastive learning for user cold-start recommendation with social variable},
  author={Xu, Xiaolong and Dong, Hongsheng and Xiang, Haolong and Hu, Xiyuan and Li, Xiaoyong and Xia, Xiaoyu and Zhang, Xuyun and Qi, Lianyong and Dou, Wanchun},
  journal={ACM Transactions on Information Systems},
  year={2025},
  publisher={ACM New York, NY}
}

@article{zeng2025cabb,
  title={Click A, Buy B: Rethinking Conversion Attribution in E-Commerce Recommendations},
  author={Zeng, Xiangyu and Jaspal, Amit and Liu, Bin and Panneeru, Goutham and Huang, Kevin and Bievre, Nicolas and Jaggi, Mohit and Maniraju, Prathap and Jain, Ankur},
  journal={arXiv preprint arXiv:2507.15113},
  year={2025}
}

@inproceedings{yao2022causalmta,
  title={Causalmta: Eliminating the user confounding bias for causal multi-touch attribution},
  author={Yao, Di and Gong, Chang and Zhang, Lei and Chen, Sheng and Bi, Jingping},
  booktitle={Proceedings of the 28th ACM SIGKDD Conference on Knowledge Discovery and Data Mining},
  pages={4342--4352},
  year={2022}
}

@article{bencina2025lidda,
  title={LiDDA: Data Driven Attribution at LinkedIn},
  author={Bencina, John and Aykutlug, Erkut and Chen, Yue and Zhang, Zerui and Sorenson, Stephanie and Tang, Shao and Wei, Changshuai},
  journal={arXiv preprint arXiv:2505.09861},
  year={2025}
}

\appendix

\section{Extended Related Work}
\label{app:related_work}

This appendix provides a detailed discussion of related works that were briefly summarized in the main text due to space constraints. We focus on three key areas: Generative Recommendation, LLM-Enhanced Recommendation, and methods specific to Dormant User Reactivation.

\subsection{Generative Recommendation}

%Generative recommendation reformulates recommendation as a sequence generation problem, where user preferences and items are modeled as discrete tokens and predictions are produced in an autoregressive manner~\cite{ wang2025generative}.
Generative recommendation models user behaviors and items as discrete tokens and predictions are produced in an autoregressive manner~\cite{ wang2025generative}.
%This paradigm departs from conventional retrieval-and-ranking pipelines by directly generating recommendation results, enabling unified modeling across multiple stages and tasks~\cite{hou2025generative}.
Recent studies introduce \emph{semantic ID} representations to bridge large item spaces with generative models.
Representative works such as TIGER~\cite{rajput2023recommender}, MTGR~\cite{han2025mtgr}, Tokenrec~\cite{qu2025tokenrec}, and OneRec~\cite{deng2025onerec} encode items into hierarchical or discrete semantic tokens, transforming recommendation into a structured token generation problem.
By significantly reducing vocabulary size and imposing semantic constraints, these methods improve scalability, generalization, and cross-task transferability.
%The semantic ID-based paradigm enables a unified framework that supports retrieval, ranking, and recommendation generation within a single model~\cite{wang2025generative}.
% Despite these advances, existing generative recommendation methods typically assume that user behavior sequences are fully observed and reliable.
% They focus on generating the next item or recommendation list conditioned on historical interactions, without explicitly modeling missing, unobserved, or latent user behaviors.
% As a result, the potential of generative recommenders to reason about incomplete user histories—particularly for under-observed users—remains largely unexplored.

\subsection{LLM-Enhanced Recommendation}
The emergence of large language models (LLMs) has opened new opportunities to alleviate data sparsity by leveraging their strong semantic understanding and generative capabilities~\cite{hua2023tutorial,lin2025can,chen2024large,wu2024survey}.
Existing LLM-based recommendation approaches can be broadly categorized into two paradigms: {(1) LLM-based representation enhancement}, which leverages LLMs to produce semantic embeddings for items or users through offline pipelines~\cite{liu2025llm-esr, liu2025llmal,cheng2025poi,ge2025llm}.  
The resulting representations are then fused with collaborative signals in downstream recommendation models.
Despite thier success, most of them integrates LLMs' outputs as side information without tightly coupling LLM reasoning with the recommendation process. 
%Despite their success, most existing methods treat LLMs as external and static feature generators, integrating their outputs as side information without tightly coupling LLM reasoning with the recommendation process. 
{(2) LLM-based data augmentation}, where LLMs are employed to generate synthetic user–item interactions or pseudo behavioral sequences~\cite{sun2025llmser, Wei2024LLMRec, liu2025filterllm, Gu2025R4ec}.
While effective in enriching training data, these methods typically offer limited controllability over generated behaviors, and are susceptible to hallucinated interactions.
Consequently, the ability of LLMs to infer and structure high-level user intent beyond observed interactions remains largely underexplored.

\subsection{Recommendation for Dormant Users
}

Dormant users represent a critical yet under-explored group, which are previously active individuals having rich historical interaction but exhibit non-conversion in a long period~\cite{kutlimuratov2022modeling}. 
%Unlike cold-start users with little historical data, dormant users possess relatively rich interaction histories. 
However, due to temporal preference drift~\cite{zhang2025cold}, these historical signals often become stale or misaligned with current user intent and platform dynamics, making effective re-engagement particularly challenging.
Existing studies on dormant user recommendation can be broadly categorized into three paradigms.
\textit{(1) Latent Preference Recovery} methods aim to infer hidden user intent from historical behaviors, using techniques such as VAE-based intent modeling~\cite{chang2023latent} or deep matrix factorization over dormant user clusters~\cite{kutlimuratov2022modeling}. 
% While effective in compressing historical signals, these approaches are limited in adapting to preference shifts that occur during prolonged inactivity.
\textit{(2) Temporal Continuity Modeling} focuses on retaining long-range interests across inactivity gaps. Representative methods, such as CSTRec~\cite{lee2025capturinguserinterestsdata} and PESO~\cite{yoo2025continuallowrankadaptersllmbased}, employ continual learning and a balance between stability and plasticity to prevent catastrophic forgetting. 
%However, these methods still rely heavily on observed behavioral sequences and struggle when recent feedback is sparse or noisy.
Recent work explores \textit{(3) LLM-based Behavioral Augmentation} for dormant users. By leveraging the world knowledge and reasoning ability of large language models, these approaches enrich sparse behavior sequences with inferred preferences or synthesized semantic context~\cite{huang2025impact}. 
%LLMs can analyze textual descriptions of historical interactions and candidate items to generate plausible preference signals, partially alleviating data sparsity.
ColdLLM~\cite{huang2024large} leverages a customized LLM-based simulator to generate user–item interactions between users and cold items, warming cold items and demonstrating LLMs’ potential in sparse settings. Beyond behavior simulation, CSRec~\cite{yang2024common} and CoLaKG~\cite{cui2025comprehending} enhance recommendation by leveraging LLMs to enrich knowledge graphs—via common-sense knowledge fusion and semantic graph embeddings, respectively, hereby strengthening relational signals under data sparsity settings.
\begin{comment}
ColdLLM~\cite{huang2024large} leverages a customized LLM-based simulator to mimic interactions between users and cold items, effectively transforming cold items into warm ones through generated behavioral patterns. Although primarily designed for cold-start items, this line of work demonstrates the potential of LLMs to simulate plausible user–item interactions with sparse feedback. Beyond behavior simulation, CSRec~\cite{yang2024common} exploits common-sense knowledge from LLMs to construct an auxiliary knowledge graph and integrates it with existing metadata graphs via knowledge fusion, enriching relational signals that are otherwise missing in sparse settings. Similarly, CoLaKG~\cite{cui2025comprehending} enhances knowledge graphs by transforming structured graph data into textual inputs for LLMs and generating semantic embeddings, thereby incorporating global graph semantics into recommendation models.
\end{comment}

Despite their promise, existing LLM-enhanced approaches only considers the standalone items individually, ignoring the user's conversion trajectory value maximization. In contrast, our work introduces a reasoning-augmented generative framework that explicitly models user intent evolution through structured functional role trajectories, enabling more principled exploration and re-engagement beyond surface-level behavior recovery.

\section{Implementation Details}
\label{app:implementation}

In this section, we provide comprehensive implementation details of \shortname{}.
We explicitly outline the {training strategies} for both the Reasoner and the Behavioral Backbone, followed by the {online deployment architecture} designed for industrial serving.
Furthermore, we detail the {data construction pipeline}, and  specific hyperparameter configurations used in our experiments(Section~\ref{app:exp_model}).

\subsection{Training Strategy}
\subsubsection{Training of the Reasoner.}
\label{sec:train_llm}
% SID扩展；多少个阶段，每个阶段解冻哪些部分
We first expand the vocabulary of the base LLM to incorporate the hierarchical SIDs, adding tokens corresponding to the three semantic levels.
To ensure robust adaptation, we adopt a progressive optimization strategy.
In the item alignment phase (Section~\ref{sec:item_align}), we exclusively unfreeze the input embedding layer and the output language modeling head (\texttt{lm\_head}). This aligns the newly added SIDs with the pre-trained semantic space while preserving the model's general linguistic knowledge.
Subsequently, for the remaining instruction tuning phases, we perform full-parameter fine-tuning.
The model is optimized using the standard SFT objective, minimizing the auto-regressive negative log-likelihood loss on the target response tokens.

\subsubsection{Training of the Behavioral Backbone.}
\label{sec:train_gen}

To capture comprehensive behavioral pattern on the our platform and avoid performance degradation caused by data sparsity, we first train the generative behavioral backbone on logs from \emph{all} users, including both dormant and active users. 
Following ~\cite{zhou2025onerecv2}, we adopt an autoregressive cross-entropy objective over the three semantic-ID tokens of each target item:
\begin{footnotesize}
\begin{equation}
\label{eq:loss_gen}
\mathcal{L}_{\text{Gen}} = -\frac{1}{L} \sum_{l=1}^{L} \log p \left( \mathrm{SID}_l(i^\mathrm{gen}) \mid \text{BOS}, \mathrm{SID}_{<l}(i^\mathrm{gen}), \text{Context} \right),
\end{equation}
\end{footnotesize}
where $L=3$ denotes the level of the hierarchical SID, $\mathrm{SID}_l(i^\mathrm{gen})$ represents the $l$-th token of the target item's SID, and $\text{BOS}$ is the begin-of-sentence token. The $\text{Context}$ compromises both static user attributes and the interaction history~\cite{zhou2025onerecv2}.

Upon completion of this general training, we further specialize the model for dormant-user recommendation via Reasoner-guided warm-start fine-tuning. 
We filter for dormant instances and augment their input context with Reasoning Guidance Features (Section~\ref{sec:synergy}).
The backbone is then continuing fine-tuned on this augmented dataset using the same objective in Eq.~\eqref{eq:loss_gen}.
% [todo: 回头检查一下这里的augment features跟前文是否对应]

% To understand the user behaviors on the our platform and align with the user pattern, we use all merchant users' samples to train the generative behavioral backbone, including both dormant and activate users. Similar to ~\cite{zhou2025onerecv2}, we adopt the cross-entropy loss across three semantic tokens as the generation loss:
%为了保证生成式行为大模型的推荐效果，我们使用所有用户的数据进行Supervised Fine-Tuning phase。与OneRec-v2~\cite{zhou2025onerecv2}一致，我们的模型trained using average generation loss of 生成的三级SID，即： 

% 在完成上述训练后，进行reasoner增强的dormant user数据热启动加训。 具体来说，选取dormant users，如Section ~\cite{sec:backbone}所述利用Reasoner的结果来augment features，并同样基于Eq.~\eqref{eq:loss_gen}进行优化。

% After the pre-train stage, we inject the Reasoner enhanced information into the behavioral model by warm-up training. As described in Section ~\cite{sec:backbone}, the Reasoner generate an item set with function role reasoning, and it serves as sequential feature into the generative model. The loss is also Eq.~\eqref{eq:loss_gen}.

%在完成上述训练后，进行reasoner增强的dormant user数据热启动加训。如Section ~\cite{sec:backbone}所述，基于Reasoner生成的item候选集构建多个新特征列，增加到原有generative model的输入context中，并同样基于Eq.~\eqref{eq:loss_gen}进行优化。As described in Section ~\cite{sec:backbone}, we construct features based on the counterfactual generated item SID set from Reasoner, and augment them to the input context of the behavior model.

\subsection{Deployment Architecture Details}
\label{app:deploy_details}

Complementing the architecture overview in Section~\ref{sec:deployment}, we here provide the specific operational workflows for the asynchronous collaborative design:

\begin{itemize}[leftmargin=*]
    \item \textbf{Near-line Reasoning Lane.} 
    The Reasoner operates in a near-line environment to handle high-latency inference tasks.
    \begin{enumerate}
        \item Initialization \& Updating: For the initial deployment, we collect logs from the past three months of successful conversions of dormant users and perform SFT following Section~\ref{sec:train_llm}. After deployment, we update the Reasoner on a weekly basis by aggregating newly collected logs of successful dormant-user conversions, predictions from the behavioral backbone, and subsequent user feedback. These signals are mixed with historical data for continual fine-tuning.
        \item Inference: On a weekly basis, we employ vLLM to perform counterfactual inference (beam size=25) for all dormant users. The derived semantic signals are structured into knowledge-guided features and cached by user ID for fast lookup at online serving.
    \end{enumerate}  
    
    \item \textbf{Online Serving Lane.} 
    The behavioral backbone serves real-time traffic as one of the retrieval channels.
    Upon an incoming request from user $u$, the backbone retrieves the pre-computed Reasoner features for $u$ from the near-line cache, integrates them with the real-time context and the latest interaction sequence, and performs next-item generation with a beam size of $1024$. 
    The generated candidates are then passed to downstream coarse-ranking and fine-ranking stages for final exposure.
    Thanks to this decoupled design, the online serving path benefits from deep functional-role CoT reasoning signals while maintaining strictly low latency (P99 $< 53$ms), ensuring a seamless user experience.
\end{itemize}

\subsection{Training Data Preparation}
\label{app:data}
% 依照Section ~\ref{sec:deployment},对于Reasoner，我们采样过去3个月的成果转化的dormant users进行SFT，包含约X用户、X item，其中交互行为包括{view, click, pruchase}三种，我们对view的样本进行了降采样，而保留所有click、pruchase样本。参考课程学习技术，在Item alignment阶段，我们首先对数据中涉及到所有item进行一轮（epoch）对齐，而后对于销量大于X的优质物品（约X万）再进行一轮对齐训练。考虑到Reasoner的训练推理开销大，交互序列中保留最新的128个优质物品。完成训练后，对X万的dormant user进行beam size为25的反事实推理，获得对应的reasoner推理候选物品集合$\{\hat{i}^{\mathrm{rea}_u}\}$，并可以构建对应的Reasoning Guidance Features。with上述X万用户的Reasoning Guidance Features，behavioral generative backbone进行热启动的训练。

Following the deployment setup, we construct the SFT dataset for the Reasoner by sampling users who transitioned from a dormant state to successful conversion within the past three months.
This large-scale dataset comprises over $22$M users and approximately $28$M items, incorporating three types of interactions: $\{ \texttt{view}, \texttt{click},$ $\texttt{purchase} \}$.
To control label imbalance and focus on high-confidence signals, we down-sample \texttt{view} events while retaining all \texttt{click} and \texttt{purchase} interactions.
Inspired by curriculum learning, we adopt a two-stage strategy in the item alignment phase (Section~\ref{sec:item_align}).
Initially, we conduct one epoch of alignment on the entire item set to establish broad semantic coverage. Subsequently, we perform an additional epoch restricted to high-quality items whose orders exceed $5$, resulting in about $9$M items.
Considering the computational overhead of the Reasoner, we truncate each interaction sequence to keep only the most recent $128$ high-quality items.

After the Reasoner is trained, we execute counterfactual inference with beam size $25$ for about $14$M dormant users, obtaining for each user a candidate set of items $\{\hat{i}^{\mathrm{rea}}_u\}$ and the corresponding \textit{Reasoning Guidance Features}.
The enriched features from these 14M users are then utilized to further fine-tune the pre-trained behavioral generative backbone.
% todo:数据包含哪些特征

\subsection{Configurations and Hyperparameters}
\label{app:exp_model}
% 【todo:确认细节】

In this section, we detail the configuration and training protocols for \shortname{} and the baseline models. First, we present the baselines information in details:
\begin{itemize}[leftmargin=*]
    \item \textbf{SASRec}~\cite{kang2018sasrec}: A state-of-the-art sequential recommendation model that employs self-attention mechanisms to capture long-term dependencies within user interaction sequences.
    \item \textbf{TIGER}~\cite{rajput2023tiger}: A pioneering generative recommendation framework. It introduces semantic IDs via RQ-VAE and formulates recommendation as a sequence-to-sequence generation task, predicting target SIDs autoregressively.
    \item \textbf{U2I} \& \textbf{I2I}: The production-tier retrieval methods currently deployed on the Kuaishou e-commerce platform, including both U2I-type and I2I-type methods. U2Is (two-tower based model) and I2Is (like Swing~\cite{yang2020large} and PDN~\cite{li2021path}) are the major online methods, serving full-scale online traffic, and providing over 80\% online exposures, clicks, and orders. % todo：介绍快手的召回概况；说明展示的两个基线的代表性（占比）serving full-scale online traffic
    
    \item \textbf{w/o Reasoner}: The base version of our proposed generative backbone, trained on the entire platform user population without the intervention of the Reasoner.
    \item \textbf{\shortname{}}: The complete framework proposed in this paper. It fine-tunes the BehBackbone by incorporating reasoning guidance features and employing the synergistic training strategy (Section~\ref{sec:synergy}).
\end{itemize}

Both the Reasoner and the Behavioral Backbone utilize a shared vocabulary derived from the in-house Semantic ID (SID) codebook deployed in Kuaishou's production environment~\cite{luo2025qarm_sid}. This codebook employs a hierarchical structure with 3 layers, containing 4,096 tokens per layer.

\paragraph{Conversion Trajectory Reasoner.} 
The Reasoner is initialized from {Qwen-8B} and fine-tuned on a cluster of 32 NVIDIA A800 GPUs. The training framework is developed upon LLaMA Factory~\cite{zheng2024llamafactory}, using a learning rate of $1.0 \times 10^{-4}$.
In the \textit{Item Semantic-ID Alignment} phase, we expand the original LLM vocabulary by adding 12,290 tokens, covering the three-level SIDs ($3 \times 4096$) and special boundary tokens (\texttt{<sid\_begin>}, \texttt{<sid\_end>}). 
Given the relatively short prompt length in this stage (P99 $<$ 130 tokens), we set the global batch size to 256 and exclusively optimize the \texttt{embed\_tokens} and \texttt{lm\_head} layers.
In the subsequent \textit{Instruction Tuning with Structured Reasoning} phase, we perform full-parameter fine-tuning. Due to the inclusion of CoT content, the prompt length increases significantly (P99 $<$ 2,900 tokens); consequently, we adjust the global batch size to 32.
During counterfactual inference, we employ beam search with a beam width of 25 and a temperature of 0.95 to generate diverse counterfactual candidates.

\paragraph{Generative Behavioral Backbone.} 
Our generative behavioral backbone is instantiated as a decoder-only architecture with approximately 0.1 billion parameters. It consists of 6 decoder layers and 4 attention heads.
We train the backbone with a global batch size of 128 and a peak learning rate of $1.0 \times 10^{-4}$, following a cosine decay schedule. The maximum context length is truncated to \texttt{128} tokens.

\paragraph{Baseline Configurations.}
For \textbf{SASRec}, we set the embedding dimension to 128, the number of self-attention blocks to 2, and the number of attention heads to 2. We use a dropout rate of 0.3 and a learning rate of $1.0 \times 10^{-3}$.
For \textbf{TIGER}, to ensure a fair comparison, we utilize the same in-house SID system described above, thereby bypassing the original RQ-VAE tokenization. The generative retrieval component is instantiated as a Transformer-based encoder-decoder architecture with 8 layers and 6 heads trained with a learning rate of $1.0 \times 10^{-4}$ and a dropout rate of 0.1.

\section{Prompt Template}
\label{app:prompt}

\subsection{Detailed prompts for Reasoner}
Table \ref{tab:item_indexing} $\sim$ \ref{tab:reasoning-sub-tasks} presents the detailed prompts for instructing Qwen-8B in multiple tasks, including SID alignment, next item prediction, and functional role reasoning.

\begin{table}[t]
  \vspace{-0pt}
  \caption{Detailed prompts for instructing Qwen-8B in Item Indexing.}
  \vspace{-0pt}
  {
    \footnotesize
    \setlength{\parindent}{0pt}
    \begin{tabular}{p{\linewidth}}
    \toprule
    \textbf{Instruction} \\
    You are a semantic indexing module responsible for mapping product metadata from the e-commerce platform to discrete Semantic ID (SID) tokens. \\
    \midrule
    \textbf{Prompt} \\
    Generate a Semantic ID (SID) token that represents the given item based on the \\ following product information: \\
    Title: \textcolor{blue!80}{\{item\_title\}} \\ 
    Price: \textcolor{blue!80}{\{item\_price\}} \\ 
    Category Path: \textcolor{blue!80}{\{category\_path\}} \\
    Output only the SID token. \\
    \midrule
    \textbf{Target}\\
    \textcolor{blue!80}{\{multi\_level\_sid\}} \\
    \bottomrule
    \end{tabular}}%
  \label{tab:item_indexing}%
\end{table}%

\begin{table*}[htbp]
  \vspace{-0pt}
  \caption{Detailed prompts for instructing Qwen-8B in Item Profiling.}
  \vspace{-0pt}
  {
    \footnotesize
    \setlength{\parindent}{0pt}
    \begin{tabular}{p{\linewidth}}
    \toprule
    \multicolumn{1}{c}{\textbf{Sub Task 1: SID $\to$ title}} \\
    \midrule
    \textbf{Instruction} \\
    You are an e-commerce index-to-text mapper. Given an internal Semantic ID (SID) token, you must generate the corresponding product title in the style commonly used on e-commerce platforms—characterized by long, keyword-rich formulations. \\
    \midrule
    \textbf{Prompt} \\
    Please generate the original e-commerce product title corresponding to the following SID.\\
    SID: \textcolor{blue!80}{\{multi\_level\_sid\}} \\
    Title: \\
    \midrule
    \textbf{Target}\\
    \textcolor{blue!80}{\{item\_title\}} \\
    \midrule
    \multicolumn{1}{c}{\textbf{Sub Task 2: SID $\to$ category path}} \\
    \midrule
    \textbf{Instruction} \\
    You are a category prediction module responsible for decoding the product category from Semantic ID (SID) tokens in an e-commerce recommendation system. \\
    \midrule
    \textbf{Prompt} \\
    Given a product Semantic ID (SID), infer its category path in the format:
    "Level-1 Category > Level-2 Category > Level-3 Category". \\ 
    SID: \textcolor{blue!80}{\{multi\_level\_sid\}} \\
    Category Path: \\ 
    \midrule
    \textbf{Target}\\
    \textcolor{blue!80}{\{category\_path\}} \\
    \bottomrule
    \end{tabular}}%
  \label{tab:item_profiling}%
\end{table*}%

\begin{table*}[htbp]
  \vspace{-0pt}
  \caption{Detailed prompts for instructing Qwen-8B in Standard behavior prediction.}
  \vspace{-0pt}
  {
    \footnotesize
    \setlength{\parindent}{0pt}
    \begin{tabular}{p{\linewidth}}
    \toprule
    \textbf{Instruction} \\
    You are a senior expert in e-commerce recommendation algorithms.\\
    Task: Given the user's personal profile and historical interaction sequence, predict the next item ID under the user's \textcolor{blue!80}{\{target\_behavior\}} action.\\
    Procedure:
    Analyze the user’s long-term interest tags and spending capacity.
    Identify recent trends and potential shifts in user intent from the interaction history.\\
    Exclude items previously interacted with in the history and recommend a novel Semantic ID (SID) that best aligns with the user’s current intent. \\
    \midrule
    \textbf{Prompt} \\
    User interactions are recorded as (behavior, item SID) pairs.
    Based on the user profile and recent interaction history, predict the next item they will interact with. \\
    Profile: \textcolor{blue!80}{\{user\_profile\_text\}} \\
    History: \textcolor{blue!80}{\{interaction\_seq\_text\}} \\
    Output only one item SID token (must be a new item not appearing in the history). \\ 
    \midrule
    \textbf{Target}\\
    \textcolor{blue!80}{\{target\_item\_token\}} \\
    \bottomrule
    \end{tabular}}%
  \label{tab:item_profiling}%
\end{table*}%

\begin{table*}[htbp]
  \vspace{-0pt}
  \caption{Detailed prompts for instructing Qwen-8B in Functional Role–guided Chain-of-Thought.}
  \vspace{-0pt}
  {
    \footnotesize
    \setlength{\parindent}{0pt}
    \begin{tabular}{p{\linewidth}}
    \toprule
    \textbf{Instruction} \\
    You are an e-commerce recommendation expert specializing in user intent understanding. Your task is to accurately predict the SID of the next item for the user’s \textcolor{blue!80}{\{target\_behavior\}} based on their profile and interaction history. 
    Please strictly follow the steps below and conduct your reasoning within tags: \\
    Profile Anchoring: Summarize the user’s core preference categories. \\
    Functional Role Analysis: Scan the interaction history to identify the functional roles of key items (e.g., high-frequency consumables, long-tail durables, novelty-seeking items).\\
    Intent Evolution Inference: Analyze the temporal evolution pattern of these roles (e.g., purchased a console (durable) → seeking a controller (complementary)), and infer the current demand gap accordingly.\\
    Decision Alignment: Based on the inferred demand gap, select the most aligned item SID.\\
    After completing your reasoning, start a new line and output only the predicted item SID. Ensure the recommendation is novel and does not duplicate any item in the historical sequence. \\
    \midrule
    \textbf{Prompt} \\
    User Profile: \textcolor{blue!80}{\{user\_profile\_text\}} \\ 
    Interaction Sequence: \textcolor{blue!80}{\{interaction\_seq\_text\}} \\
    Please reason: What functional need is reflected by the key items the user has recently engaged with? Based on this need, what is the most reasonable next item? \\ 
    Output your reasoning process followed by the SID. \\
    \midrule
    \textbf{Target}\\
    <think>\textcolor{blue!80}{\{think\_content\}}</think>\textcolor{blue!80}{\{target\_item\_token\}} \\
    \bottomrule
    \end{tabular}}%
  \label{tab:item_profiling}%
\end{table*}%

\begin{table*}[htbp]
  \vspace{-0pt}
  \caption{Detailed prompts for instructing Qwen-8B in Stepwise reasoning sub-tasks.}
  \vspace{-0pt}
  {
    \footnotesize
    \setlength{\parindent}{0pt}
    \begin{tabular}{p{\linewidth}}
    \toprule
    \multicolumn{1}{c}{\textbf{Sub Task 1: Key Items Extraction}} \\
    \midrule
    \textbf{Instruction} \\
    You are an e-commerce data analyst. Analyze the user profile and interaction history to extract the key items most informative for predicting the user’s next \textcolor{blue!80}{\{target\_behavior\}}. \\
    Selection criteria: Retain items that are either strongly aligned with the user’s core preferences or indicative of recent shifts in demand. \\
    Output format: List the SIDs of the selected items in chronological order of interaction, one SID per line. \\
    \midrule
    \textbf{Prompt} \\
    User Profile: \textcolor{blue!80}{\{user\_profile\_text\}}
    Interaction History: \textcolor{blue!80}{\{interaction\_seq\_text\}}
    Please filter the interaction sequence based on the user profile to identify Key Items:\\
    \midrule
    \textbf{Target}\\
    \textcolor{blue!80}{\{key\_items\_list\_str\}} \\
    \midrule
    \multicolumn{1}{c}{\textbf{Sub Task 2: Functional Role Interpretation.}} \\
    \midrule
    \textbf{Instruction} \\
    You are a recommendation expert on an e-commerce platform. Analyze the functional role of the given key item.
    Incorporate the item’s intrinsic attributes—sales volume, price, and durability—together with the user’s core preference categories to determine its role in the user’s decision-making path.
    Role analysis dimensions: \\
    Sales frequency: (high\_freq, mid\_freq, low\_freq),\\
    Price level: (high\_price, mid\_price, low\_price),\\
    Durability: (durable\_cate, freq\_cate), \\
    and the relationship to the user’s core preference categories (e.g., category substitution, complementary pairing, trend-following within peer group).
    Output format:
    Attributes: {sales/price/durability}, relationship to user preference category is \textcolor{blue!80}{\{relationship type\}}.\\
    \midrule
    \textbf{Prompt} \\
    User Information: \textcolor{blue!80}{\{user\_profile\_text\}} \\
    Interaction Sequence: \textcolor{blue!80}{\{interaction\_seq\_text\}}\\
    Key Items: \textcolor{blue!80}{\{key\_items\_list\_str\}} \\
    Infer the functional roles provided by these key items: \\
    \midrule
    \textbf{Target}\\
    \textcolor{blue!80}{\{functional\_role\}} \\
    \midrule
    \multicolumn{1}{c}{\textbf{Sub Task 3: Joint Reasoning \& Intent Modeling.}} \\
    \midrule
    \textbf{Instruction} \\
    As an expert in e-commerce intent inference, your task is to analyze the user's interaction history and build a logical chain leading to the user's next \textcolor{blue!80}{\{target\_behavior\}} decision. Here's how to proceed with the analysis: \\
    Clue Extraction: From the interaction sequence, filter out the noise and focus on key items that reflect the core intentions of the user. \\
    Role Decoding: Based on the user profile, analyze the functional roles these key items play in satisfying the user's needs or desires. \\
    Intent Modeling: Predict what the functional role of the item for the user's next \textcolor{blue!80}{\{target\_behavior\}} might be. \\
    Given that we don't have specific data such as \textcolor{blue!80}{\{user\_profile\_text\}}, \textcolor{blue!80}{\{interaction\_seq\_text\}}, and \textcolor{blue!80}{\{key\_items\_list\_str\}}, I'll outline a template for the structured analytical report: \\
    \midrule
    \textbf{Prompt} \\
    User Profile: \textcolor{blue!80}{\{user\_profile\_text\}} \\
    Interaction History: \textcolor{blue!80}{\{interaction\_seq\_text\}} \\
    Reasoning Task: Identify key items, analyze their functional roles, and ultimately predict the functional role of the next item. \\
    \midrule
    \textbf{Target}\\
    \textcolor{blue!80}{\{think\_content\}} \\
    \bottomrule
    \end{tabular}}%
  \label{tab:reasoning-sub-tasks}%
\end{table*}%

\end{document}